\documentclass[journal]{IEEEtran}
\usepackage{tfrupee}  
\usepackage{amsmath}
\usepackage{amsfonts} 
\usepackage{lipsum}
\usepackage{resizegather}
\usepackage{textcomp}
\usepackage{multicol}
\usepackage{mathtools}
\usepackage{multirow}
\usepackage{tabularx,booktabs}
\usepackage{tabularx}
\graphicspath{./GTIBDRPicx/}
\usepackage{cuted}
\usepackage{tcolorbox}
\usepackage{csquotes}
\usepackage{subcaption}
\usepackage{nomencl}
\usepackage{etoolbox}
\usepackage{multibib}
\usepackage[ruled, linesnumbered]{algorithm2e}
\usepackage{listings}
\SetKwInput{KwInput}{Input}
\SetKwInput{KwOutput}{Output}
\usepackage{enumitem}
\usepackage{xcolor}
\usepackage{color,soul}

\usepackage{cite}
\ifCLASSINFOpdf
   \usepackage{graphicx}
   \usepackage{subcaption}
   \graphicspath{{../GTIBDRPicx/}{../eps/}}
   \DeclareGraphicsExtensions{.pdf,.jpeg,.png,.eps}
\else
\fi
\usepackage{algorithmic}
\usepackage{url}
\begin{document}
\title{A Bi-level Decision Framework for Incentive-Based Demand Response in Distribution Systems}
\author{Vipin~Chandra~Pandey*,
        Nikhil~Gupta, \emph{Senior Member, IEEE}, Khaleequr Rehman Niazi, \emph{Senior Member, IEEE},~Anil~Swarnkar, \emph{Senior Member, IEEE},~Tanuj Rawat,~and~Charalambos~Konstantinou, \emph{Senior Member, IEEE}
        \thanks{Vipin Chandra Pandey and Charalambos Konstantinou are with the Computer, Electrical and Mathematical Sciences and Engineering (CEMSE) Division, King Abdullah University of Science and Technology (KAUST), Thuwal 23955-6900, Saudi Arabia (e-mail: vipinntl@gmail.com).}
        \thanks{Nikhil Gupta, Khaleequr Rehman Niazi, Anil Swarnkar, and Tanuj Rawat are with the Electrical Engineering Department, Malaviya National Institute of Technology (MNIT), Jaipur 302017, India.}
        \thanks{*Work done primarily while the author was with MNIT Jaipur.}
        }
\IEEEaftertitletext{\vspace{-2.5\baselineskip}}
\maketitle

\begin{abstract}
In a growing retail electricity market, demand response (DR) is becoming an integral part of the system to enhance economic and operational performances. This is rendered as incentive-based DR (IBDR) in the proposed study. It presents a bi-level decision framework under the ambit of multiple demand response providers (DRPs) in the retail competition. It is formulated as a multi-leader-multi-follower game, where multiple DRPs, as the DR stakeholders, are strategically interacting to optimize load serving entity cost at the upper level, and individual DRP as the aggregated customers is optimizing its cost at the lower level. The strategic behavior of DRPs is modeled in a game-theoretic framework using a generalized Stackelberg game. Further, the existence and uniqueness of the game are validated using variational inequalities. It is presented as a nonlinear problem to consider AC network constraints. An equilibrium problem with equilibrium constraints is used as a mathematical program to model the multi-leader-multi-follower, bi-level problem, which is simultaneously solved for all DRPs. The diagonalization method is employed to solve the problem. The detailed numerical analyses are conducted on IEEE 33-bus test and Indian-108 bus distribution systems to demonstrate the applicability and scalability of the proposed model and the suggested method.
\end{abstract}

\begin{IEEEkeywords}
Demand response, bi-level problem, equilibrium problem with equilibrium constraints, generalized Stackelberg game.
\end{IEEEkeywords}

\makenomenclature
\section*{Nomenclature}
\addcontentsline{toc}{section}{Nomenclature}
\begin{IEEEdescription}[\IEEEusemathlabelsep\IEEEsetlabelwidth{~~~~~~~~~~~~~}]
	\item[\textit{\textbf{Indices and Sets}}]
	\item[$t$, $d$, $n$]  Indices for time, DRPs and bus.
	\item[${\Omega_\textbf{T}}$, ${\Omega _\textbf{D}}$,  ${\Omega _\textbf{N}}$] Sets for time, DRPs/customer classes and bus.
    \item[$n, m$, ${\Omega _\textbf{L}}$] Index and set of lines.
			\item[\textit{\textbf{Parameters and Constants}}]
	\item[$\rho _t^{RTP}$]  LSE purchasing rate from wholesale market at time \textit{t}.
	\item[$\rho _t^{FR}$] LSE selling retail/flat rate at time $t$. 
	\item[$\rho _{d,t}^{FR}$] LSE selling Electricity rate to customer class/DRP $d$ at time $t$.
	\item[$P_{n,t}^{F}$]  Flexible active load of $n_\text{th}$ bus at time $t$.  
	\item[$Q_{n,t}^{F}$]  Flexible reactive load of $n_\text{th}$ bus at period $t$.  
	\item[$P_{n,t}^{IF}$]  Inflexible active load of $n_\text{th}$ bus at period $t$.  
	\item[$Q_{n,t}^{IF}$]  Inflexible reactive load of $n_\text{th}$ bus at period $t$.  
	\item[${P_{L,n,t}},{Q_{L,n,t}}$] Active and reactive load on $n_\text{th}$ bus at time $t$.
	\item[$P_t^{max}$] Permissible maximum flexible demand at time $t$. 
	\item[$\kappa_d$] Subsidy/Overcharging factor of class \textit{d}.
	\item[$P_{d,t,o}^{F}$] Flexible demand under DRP $d$ before DR at time $t$.
		\item[\textit{\textbf{Functions and Variables}}]
		\item[$\rho _{t}^{INC}$] Incentive rate offers at time $t$.
      \item[$\rho _{d,t}^{INC}$] Incentive rate offers to DRP $d$ at time $t$.
      \item[$P_{d,t}^F$] Flexible demand under DRP $d$ at time $t$.
      \item[${P_{G,n,t}}$]  Active power generation  on bus $n$ at period $t$.
       \item[${Q_{G,n,t}}$] Reactive power generation on bus $n$ at time $t$.
       	\item[$P_{Loss,t}$] Active power loss at time $t$.
      \item[${e_n},{f_n}$] Rectangular coordinates of voltage on bus $n$.
      \item[${V_n}$] Absolute voltage of $n_\text{th}$ bus.
      \item[${\widehat S_{nm,t}}$] Mean total power flow between line connected to bus $n$ and $m$.
      \item[${P_{L,t}}$]  Total active power load at period $t$.
    	\item[\textit{\textbf{Symbols}}]
    \item[ $\underbar{(.)}, \overline{( . )}$ ]  Lower and upper value of $(.)$.
\end{IEEEdescription}

\vspace{-1mm}
\section{Introduction}
Demand response (DR) is considered a progressive development for   power systems. It strengthens the system’s resiliency and security by mitigating sudden peak demand rise, network congestion, variability of renewable generations, etc~\cite{schweppe2013spot}. This, in turn, helps to scale down the monopolies of generator companies, price volatility, deferment of distribution network expansion, and accommodation of renewable resources~\cite{schweppe2013spot}. DR has been part of wide applications in the wholesale electricity market~\cite{kirschen2000factoring} and has also been evolving as a dedicated DR-based market model such as demand response exchange~\cite{nguyen2010pool}. It offers financial incentives or dynamic prices for shaving or adjustment of demand during peak period~\cite{qdr2006benefits}. Further, it has also been proliferating in distribution systems (DS) with the advent of smart grid technologies in recent times~\cite{kirschen2000factoring}. This empowers the small retail to large customers for DR expansion~\cite{boroumand2015electricity,song2017price}. These developments lead to the emergence of multiple resource players, which act as the stakeholders and are not owned by DS~\cite{algarni2009generic,haider2020toward}. Under such circumstances, the behavior of the different stakeholders can be viewed as a strategic interaction. 

The strategic interactions have been vital to illustrate the competition among the different entities such as generators, suppliers, customers, demand response providers (DRPs), etc., in the electricity market~\cite{bautista2007formulation,hobbs2000strategic,cardell1997market}. In such problems, an individual’s action is influenced by rival players' actions~\cite{su2004sequential} and typically competes non-cooperatively. This leads to multiple leader’s decision-making problem, which is delineated as a multi-leader-multi-follower (MLMF/MLF) game in the game theory~\cite{bautista2007formulation,leyffer2010solving}. Its equivalent mathematical form is termed as an equilibrium problems with equilibrium constraints (EPEC), which is an amalgamation of several mathematical problems with equilibrium constraints (MPECs)~\mbox{\cite{su2004sequential}.} MPEC is the simplest form of Stackelberg game (SG), which is described through Karush-Kuhn Tucker (KKT) conditions to obtain the stationary conditions~\cite{de2003stackelberg}. Similarly, the EPEC model is formed by combining the equivalent KKT conditions of all the leaders~\cite{su2004sequential}. Further, EPEC applications have been widely applied in the strategic gaming analysis of deregulated electricity markets~\cite{su2004sequential, bautista2007formulation}. 

It has also been extended in DS in the literature~\cite{algarni2009generic,haghighat2012bilevel,jimenez2007competitive,boroumand2015electricity}. In~\cite{algarni2009generic}, the authors suggested a two-stage optimal decision framework to optimize day-ahead and real-time cost through the distributed generations (DGs) and interruptible loads in DS. But it only considers spot price as a strategic interaction. It is better demonstrated in~\cite{haghighat2012bilevel}, where the authors evinced a strategic interaction among the multiple distribution company (discos) via DGs, considering distribution as well as transmission side. A DG integration mechanism is suggested to include DGs as the competing power producers in the wholesale market for the competition~\cite{jimenez2007competitive}. In~\cite{lopez2010optimal}, an optimal contract pricing scheme for dispatchable DGs is formulated as a bi-level problem, where DGs owners maximize their profits in the upper-level and disco minimizes its operating cost at the lower-level. These strategic interactions have been limited to the asset-light suppliers, such as DGs and energy storage systems (ESSs) in DS~\cite{boroumand2015electricity}. However, DR also emerges as a viable option with the increase in the customer’s active participation due to the advanced metering infrastructure lately~\cite{faruqui2012ethics,haider2020toward}. Though, individual customer's participation is insufficient to induce demand reduction in DR. This encourages the customers to participate in DR through DRPs/DR aggregators/load aggregators\mbox{\cite{li2016dynamic,sarker2014optimal,wang2019ensuring}}. These provide an aggregated demand through the aggregated customers. Further, it establishes the coordination between the load serving entity (LSE) and customers in DR.  

This paper focuses on a bi-level IBDR framework comprised of LSE and multiple DRPs under the retail market in DS. It models DRPs as the DR stakeholders in the retail market within DS. It devises a strategic interaction among the multiple DRPs for incentive rate valuation. Many applications of DRPs have been employed in literature. A decentralized framework consists of DR aggregators and the customers are envisioned to optimizes their objectives via price and incentives-based DR for mitigating the overloading of distribution networks~\cite{sarker2014optimal}. The multiple DR aggregators optimize their profits by offering incentives to the customers. Further, the customers also minimize their costs by altering their demands in bi-level problem~\cite{wang2019ensuring}. In~\cite{hu2016framework}, the authors proposed a DR aggregation mechanism for residential customers by offering financial incentives for minimizing LSE operating cost. It also accounts the degree of comfort levels via a scaling approach. An IBDR based retail competition in the domain of multiple retailers consisting of DGs and ESS are proposed to optimize its payoff~\cite{ghazvini2015multi}. It lowers financial loss and defers the future capacity charge by employing its own asset-light resources in distribution networks. This work is further extended using a stochastic programming to incorporate uncertainty associated with renewable generation and loads~\cite{ghazvini2015incentive}. A coupon-incentive based DR framework is suggested to optimize the operational cost during peak periods under the ambit of multiple load aggregators to reflect the diversity in their operational costs~\cite{li2016dynamic}. It is an extension of~\cite{zhong2012coupon}, where the customers are offered a coupon-incentive at a flat rate (FR). 

Multiple DRPs/DR aggregators/load aggregators have been widely included to improve the system's operational and technical objectives in DS. However, their valuations in a competitive environment have rarely been part of the existing studies. This overlooks competitiveness for determining incentive price and its subsequent impact on the system's objectives. It is investigated for single DRP in~\cite{sarker2014optimal}. In~\cite{wang2019ensuring}, a strategic interaction is established between LSE and DR aggregators, but it is not considered among DR aggregators. A similar work is also considered in~\cite{hu2016framework}, where the multiple residential DR aggregators interact with LSE only. In~\cite{ghazvini2015multi}, the competition among the multiple retailers is modeled using DGs and ESS, but not with DRP. Similarly in~\cite{li2016dynamic}, DRP's strategic behavior is not considered, though incentive rates are weighted using a Shapley value to show their influence on the system. Moreover, IBDR decision based frameworks in DS usually have been designed using the straightforward assumptions such as without power flow\mbox{\cite{hu2016framework,ghazvini2015multi,li2016dynamic}} or linear AC power flow\mbox{\cite{sarker2014optimal}}, or DC power flow\mbox{\cite{wang2019ensuring}}. This ignores the attributes of the practical operation, where non-linearity and non-convexity exist in the objectives and constraints. In addition, power flow's control or state variables such as reactive power, voltage, active \& reactive power losses, etc., are essential to exhibit a pragmatic operation in distribution networks. Besides, IBDR applications have been limited to specific customer classes such as residential and industrial\mbox{\cite{sarker2014optimal,yu2017incentive,pandey2021hierarchical}}.  This shows a shortfall of inclusiveness in the existing studies. Hence, these aspects are contributed as the main motivation of the study. 

Furthermore, the proposed problem is modeled as a MLMF game. Such problems arise when each \mbox{player’s} strategy set depends upon the rival \mbox{players'} strategy sets\mbox{\cite{facchinei2007generalized}}, and when their strategies are coupled by a common coupling constraint. It is illustrated through a Generalized Stackelberg game (GSG) rather then the classical SG\mbox{\cite{pang2005quasi}}. A GSG is illustrated between power grid and plug-in electric vehicles (PEVs) for strategically determining price for PEVs with the fixed capacity constraints~\cite{tushar2012economics}. A similar work is also presented in~\cite{li2018interactive}, where the GSG between power grid and electric vehicles is suggested for interactive decision-making process. The MLMF game is solved using the diagonalization method based on Jacobi or Gauss-Seidel approach~\cite{bautista2007formulation,hobbs2000strategic,cardell1997market}. It involves an iterative process, in which a single MPEC problem is solved at a time and the process is repeated until the solution of each MPEC reaches a fixed point. 

This paper presents a IBDR framework under the ambit of multiple DRPs in a competitive environment to imitate the retail electricity market in DS. The multiple DRPs are assumed to act as light asset resources or stakeholders. The problem is formulated as a nonlinear problem (NLP) considering AC network constraints. It is described using a bi-level method, where the LSE in conjunction with multiple DRPs operates at the upper level (UL) and DRPs as the large customers or aggregators at the lower level (LL). It is illustrated as the MLMF game. Further, the strategic behavior of DRPs is exemplified in a game-theoretic framework using GSG. The equivalent mathematical problem is modeled as the EPEC and is solved using the diagonalization method~\cite{su2004sequential,leyffer2010solving}. 
The main contributions of the paper are summarized as follows:
\begin{itemize}
	\item A bi-level economic decision framework considering multiple DRPs is proposed under retail market structure using IBDR in DS. The multiple DRPs are the retail stakeholders for inducing demand flexibility during peak periods. The framework is described as a nonlinear optimization problem considering AC network constraints.
	\item The formulated bi-level problem is designed into a MLMF game from the viewpoint of the retail stakeholders (i.e., DRPs). Further, their interactions are illustrated in a  non-cooperative game-theoretic framework using GSG. In UL, the LSE optimizes its cost through coordination with DRPs, while also taking the responses of other DRPs into account. In LL, individual DRP as a large customer or aggregator optimizes its objective.  
	\item The proposed bi-level IBDR problem is mathematically formulated using the EPEC program, which simultaneously solves the multiple MPECs to find the equilibrium. It augments the LL problem into the UL using KKT conditions, and obtains its equivalent NLP formulation by combining with the stationary conditions of the UL problem. The diagonalization method is employed to solve the problem.
\end{itemize}

The remainder of the paper is outlined as follows: a bi-level IBDR problem formulation and its GSG with the deduction of existence and uniqueness  conditions are described in Section II. The solution method using NLP formulation and a pseudo code of the solution method are presented in Section III. Numerical results are discussed in Section IV and conclusions are presented in Section V.

\vspace{-1mm}
\section{Proposed IBDR Framework}

\begin{figure}[t]
	\centering 
\includegraphics[width=0.9\columnwidth]{{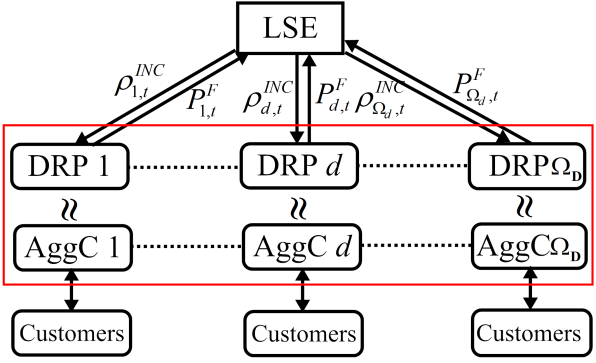}}
	\caption{Schematic diagram of the proposed IBDR framework.}
        \vspace{-3mm}
	\label{LSE_DRP}
\end{figure}

The proposed IBDR framework consists of LSE and DRPs in DS. Distribution system operator (DSO) oversees the operations of LSE and DRPs for maintaining the network security using a security-constrained AC optimal power flow (OPF). Furthermore, DSO acts as neutral market facilitators for DRPs and provides the demand flexibility, through interaction between LSE and DRPs in DS\mbox{\cite{anisie2019future, apostolopoulou2016interface,ahlstrom2015evolution, jadhav2022emergence, haider2020toward,bai2017distribution}}. The LSE is a retail stakeholder in DS. It procures the power from the wholesale market at real-time price (RTP) and supplies to the end-customers at flat rate\mbox{\cite{li2016dynamic,zhong2012coupon, sattarpour2018load, hatami2009optimal}}. The DRP is also the retail stakeholder. It operates as an independent entity and light-asset resource for providing the demand flexibility\mbox{\cite{boroumand2015electricity}}. In the proposed formulation, DRPs are assumed to be active participants and is not owned by the LSE in DS. It is defined to have primarily two functions, first for interaction between LSE and customers, and second for inducing the aggregated flexible demand from the customers. The aggregated demand is obtained using the cluster of customers, which is represented through the aggregated customers (AggCs)\mbox{\cite{li2016dynamic}}. These AggCs directly pass the aggregated flexible demand to DRPs based on the received incentives offers. This makes AggCs equivalent to DRPs from the customers' point of view as shown in Fig.~\mbox{\ref{LSE_DRP}}.  

\begin{figure}[b]
	\centering 
	\vspace{-2mm}
        \includegraphics[width=\columnwidth]{{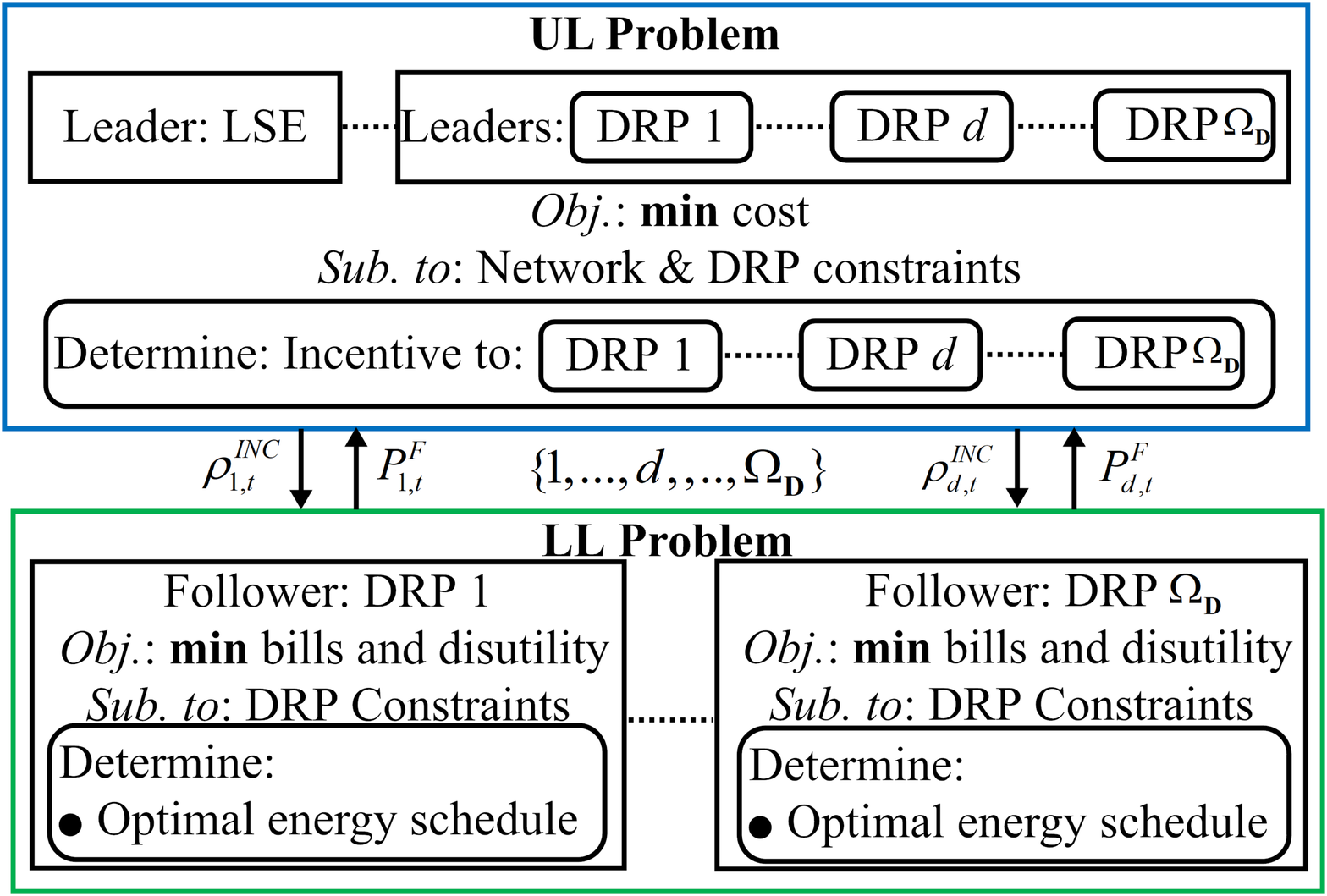}}
	\caption{MLMF Structure of the proposed IBDR framework.}
	\label{IBDR_arch}
\end{figure}

The proposed IBDR problem is devised in a non-cooperative environment. It considers interaction between LSE and DRPs as well as among the DRPs to demonstrate the competition. LSE offers incentives to DRPs using IBDR for optimizing its operating costs against high price fluctuations during peak periods\mbox{\cite{li2016dynamic,zhong2012coupon}}. This, in response, induces demand curtailment from the AggCs as received by DRPs. Since, DRPs are devised to compete with each other for incentive valuation. This makes DRP a leader in terms of game theory, whose decision or control variable (i.e., incentive rate) can influence the outcome of other DRPs and LSE\mbox{\cite{su2004sequential,facchinei2007generalized}}. Therefore, the strategic interactions among the DRPs and between LSE and DRPs are formulated using the MLMF game\mbox{\cite{su2004sequential,haghighat2012bilevel}}. It is formulated in a bi-level framework. The structure of the proposed problem using MLMF game is illustrated in Fig.~\mbox{\ref{IBDR_arch}}.

In UL, LSE and multiple DRPs interact to optimize its overall costs. It determines the incentives for DRPs and is obtained using a non-cooperative behavior of DRPs (i.e. DRPs compete independently). In LL, DRPs as the AggCs respond to offered incentive by inducing DR\mbox{\cite{wang2019ensuring}}. It is worth noting that DRPs' roles at the UL and LL are different from reference point of view. DRPs at the UL indicate the DRP's incentive cost paid by LSE, whereas DRPs at the LL represent the AggCs/large customers' objective functions. This gives incentive rate at UL, and induced demand at LL. The problem formulation of the proposed framework is described in the following section. The dual variables associated with each constraint are written after each colon.

\vspace{-2mm}
\subsection{Upper Level Problem: LSE Cost}
The objective of LSE is to optimize its operating cost. It offers incentives to the DRPs for inducing DR during the peak period. The problem is formulated for day-ahead operation considering AC network constraints and is expressed as follows\mbox{\cite{algarni2009generic,negash2014allocating,hurley2013demand}}:
\begin{multline}
	{Z_{LSE}} = \sum\limits_{t \in {\Omega _T}} {\left\{ {\rho _t^{RTP}({P_{L,t}+P_{Loss,t}})} \right.}  - \rho _t^{FR}\left( {\sum\limits_{n \in {\Omega _N}} {{P_{L,n,t}}} } \right.\\ 
	\left. {\left. { - \sum\limits_{d \in {\Omega _D}} {P_{d,t}^F} } \right) + \sum\limits_{d \in {\Omega _D}} {\rho _{d,t}^{INC}P_{d,t}^F} } \right\}
\end{multline}
\begin{gather}{\label{eq.1}}
{\rm{subject ~to}} ~~\underline \rho  _{d,t}^{INC} \le \rho _{d,t}^{INC} \le \overline \rho  _{d,t}^{INC}{\rm{:}}{\underline \mu  _{\rho ,d,t}}{\rm{, }}{\overline \mu  _{\rho ,d,t}}{\rm{  }}\\
\rho _{d,t}^{INC} \ge \rho _t^{FR}:{\mu _{\rho ,d,t}}\\
{P_{L,n,t}} = P_{n,t}^{IF} + P_{n,t}^F\\
{Q_{L,n,t}} = Q_{n,t}^{IF} + Q_{n,t}^F\\
\begin{array}{l}
	{P_{G,n,t}} - {P_{L,n,t}} + {\vartheta _{n,d}}P_{d,n,t}^F = \sum\limits_m {\left\{ {{e_{n,t}}({g_{nm}}{e_{m,t}} - {b_{nm}}{f_{m,t}})} \right.} \\
	\left. { + {f_{n,t}}({b_{nm}}{e_{m,t}} + {g_{nm}}{f_{m,t}})} \right\}:{\lambda _{p,n,t}}{\rm{,  }}\forall n \in {\Omega _\textbf{N}}
\end{array}\\
\begin{array}{l}
	{Q_{G,n,t}} - {Q_{L,n,t}} + {\vartheta _{n,d}}Q_{d,n,t}^F = \sum\limits_m {\left\{ {{f_{n,t}}({g_{nm}}{e_{m,t}} - {b_{nm}}{f_{m,t}})} \right.} \\
	\left. { - {e_{n,t}}({b_{nm}}{e_{m,t}} + {g_{nm}}{f_{m,t}})} \right\}:{\lambda _{q,n,t}}{\rm{,}},{\rm{ }}\forall n \in {\Omega _\textbf{N}}
\end{array}\\
{\left| {{{\underline V }_n}} \right|^2} \le V_{n,t}^2({e_n},{f_n}) \le {\left| {{{\overline V }_n}} \right|^2}:{\underline \pi  _{v,n,t}},{\overline \pi  _{v,n,t}}{\rm{      }}\forall n \in {\Omega _\textbf{N}}
\end{gather}
\begin{gather}
{\left| {{{\widehat S}_{nm,t}}(e,f)} \right|^2} \le {\left| {{{\widehat S}_{nm,t,\max }}} \right|^2}{\rm{:}}{\pi _{s,nm,t}}{\rm{      }}\forall (n,m) \in {\Omega _\textbf{L}}\\
{\underline P _{G,n,t}} \le {P_{G,n,t}} \le {\overline P _{G,n,t}}:{\underline \pi  _{p,n,t}},{\overline \pi  _{p,n,t}}\\
{\underline Q _{G,n,t}} \le {Q_{G,n,t}} \le {\overline Q _{G,n,t}}:{\underline \pi  _{q,n,t}},{\overline \pi  _{q,n,t}}
\end{gather}

Eq. (1) consists of three terms, first term is purchasing cost from the wholesale market, second term is the revenue cost (selling cost) to the customers, and third term is cost paid to DRPs for inducing DR. This defines the LSE's overall operating cost as the sum of purchasing cost and incentive (DRPs) costs against the selling (revenue) cost for the transacted power. Constraint (2) describes the offered incentives rates' upper and lower limits. Eq. (3) defines incentive rate constraint. Constraints (2)-(3) describe the conditions that a customer would not exhibit the demand reduction below a minimum price, and LSE would not pay to DRPs more than maximum price for the demand reduction in DR\mbox{\cite{li2016dynamic}}. The expressions (4)-(5) denote the total composition of active and reactive demands. It has been partitioned into flexible and inflexible demand, as the customers has limited flexible demands. Therefore, it is taken as the fraction of the total demand for DR illustration. The active and reactive power flows constraints are indicated in (6)-(7). It is described in rectangular coordinates due to easiness in the formulation and computation as discussed in the existing studies~\cite{bautista2007formulation,haghighat2012bilevel}. DRP's locations at particular bus in DS is defined through a binary state $\vartheta$ with ${\vartheta _{n,d}} = \{ 0,1\} {\rm{  }}~~\forall d \in {\Omega _\textbf{D}}$ in (6)-(7). It is ${\vartheta _{n,d}} = 1 {\rm{  }}$ for DRPs and ${\vartheta _{n,d}} = 0 {\rm{  }}$ for no DRPs locations. Moreover, these locations are the predetermined. Voltage bounds, and power flow limits in the lines are defined in constraints (8)-(9). The active and reactive power's bounds are designated in constraints (10)-(11). 

\vspace{-2mm}
\subsection{Lower Level Problem: (DRP \textit{d}) Model }
In LL, individual DRP induces demand flexibility based on the received incentives rates offers. Since DRP represents the cluster of the participated customers to increase its market demand in DR~\cite{li2016dynamic}. Hence, DRPs stand for aggregated or large customers at LL. This makes DRPs to also include the customers’ aspects in an individual objective at LL. Moreover, a discrete set of customer classes viz., residential (R), large industrial (LI), medium industrial (MI), agricultural (A), and commercial (C) are considered\mbox{\cite{pandey2021hierarchical}}. It considers their typical load patterns in IBDR. It is assumed that a single DRP oversees a particular customer class. This makes incentive rates and selling prices to vary uniquely. It is on account of co-existing demand diversity, activity usage, societal aspects, cost of distributing electricity, and cross-subsidy among the customer classes\mbox{\cite{faruqui2012ethics}}. It is illustrated through a factor \mbox{$\kappa_d$}, termed as subsidy (negative value)/overcharging (positive value) factor (SC/F). This makes some customer classes receive subsidized/lower rates at the expense of other customer classes. It is defined using the following relation~\cite{pandey2021hierarchical}. 
\begin{equation}
\rho _{d,t}^{FR/INC} = \rho _t^{FR/INC} \times (1 + {\kappa _d})~~~\forall d \in \Omega_\textbf{D}
\end{equation}

Eq. (12) describes the price or incentive received at the customers/DRP from the LSE level. It indicates that each DRP/customer will receive distinct price based on their SC/F. Based on received price at the LL, DRP's objective is formulated as the combination of the customer bill and dis-utility function. It is represented with DRP constraints as follows: 
\begin{gather}
\mathop {\min } {Z_{DRP,d}} = \sum\limits_{t \in {\Omega _\textbf{T}}} {\left\{ {{w_{d,1}}[(P_{d,t,o}^F - P_{d,t}^F)\rho _{d,t}^{FR} - P_{d,t}^F\rho _{d,t}^{INC}] + {w_{d,2}}{U_D}(P_{d,t}^F)} \right\}} \\
{\rm{subject ~~to:~~}}\underline P _{d,t}^F \le P_{d,t}^F \le \overline P _{d,t}^F{\rm{:}}{\underline \nu  _{{P^{F}},d,t}}{\rm{,}}{\overline \nu  _{{P^F},d,t}}{\rm{,   }}
\end{gather}
\begin{gather}
\sum\limits_{d \in {\Omega _D}} {P_{d,t}^F}  \le P_{t}^{\max }:{\nu _{{P^F},d}}~~~\forall d \in \Omega_\textbf{D},~~\forall t\\
P_{d,t}^F = \Pi (\rho _{d,t}^{INC})\\
\Pi (\rho _{d,t}^{INC}) = \frac{{{\alpha _d}\overline P _{d,t}^F}}{{(\overline \rho  _{d,t}^{INC} - \underline \rho  _{d,t}^{INC})}} \times \left( {{\chi _d}\rho _{d,t}^{INC} - \underline \rho  _{d,t}^{INC}} \right)
\end{gather}

Eq. (13) consists of two terms, first is saving in the customer bills and second is the dis-utility function. The first term is bill saving on account of demand reduction and incentive cost. The second term is accounted for the customer’s discomfort due to demand curtailment. Since, the customer lowers its bill by participating in IBDR, but it also increases its discomfort due to the demand curtailment. Therefore, both functions are simultaneously being optimized using a weighted sum approach\mbox{\cite{zadeh1963optimality}}. It assigns the weight to the individual objective to indicate their importance in the aggregated objective function with the condition of $\sum\nolimits_{m = 1}^2 {{w_{d,m}}}  = 1;\forall d,{w_{d,m}} \in (0,1)$\mbox{\cite{zadeh1963optimality}}. Constraint (14) defines the upper and lower flexible demand bounds under DRP $d$ on account of limited demand flexibility. The total demand contributed from all the DRPs in DR is kept below at a fixed demand as defined in the constraint (15). It is called a common-coupling or a shared constraint due to presence of all DRPs' induced demand terms\mbox{\cite{facchinei2007generalized}}. It signifies the contribution from all DRPs, which depends upon their functions (17). Moreover, it maintains a proportion of demand supplied by all large customers or DRPs. The incentive rate offers are assumed to be dynamic to increase DR attractiveness among the different customer class as defined in (16)~\cite{zhong2012coupon}. This makes induced demand dependent on incentive rate offers and types of the customer class. A linearly increasing function is illustrated in (17). $\alpha_d$ and $\chi _d$ are the coefficient parameters of incentive function for class $d$. These parameters describe the variation in incentive rate offers among the different customer classes~\cite{zhong2012coupon,mahmoudi2017bottom}. The dis-utility function of the aggregated customers is defined as follows~\cite{yu2017incentive}:  
\begin{equation}
{U_D}(P_{d,t}^F) = \frac{1}{2}{\theta _d}{(P_{d,t}^F)^2} + {\gamma _d}P_{d,t}^F,{\rm{    }}{\theta _d} > 0,{\gamma _d} > 0
\end{equation}    

Eq. (18) describes the aggregated customers' dissatisfaction/dis-utility cost. Further, its coefficient parameters $\theta_d$ and $\gamma_d$  denote the customer's attributes towards DR participation~\cite{yu2017incentive}. It measures the dissatisfaction on sacrificing or giving up services.

\vspace{-2mm}
\subsection{Equilibrium Model }
The formulated problem reveals a strategic interaction of DRPs over incentive rates for the possible demand curtailment. This is equivalent to Bertrand or price competition. However, Bertrand competition does not consider the capacity constraint~\cite{allen1986bertrand}. Hence, its extended model, the Bertrand-Edgeworth model also called price-demand competition, is deliberated~\cite{maskin1986existence}. It imposes capacity constraint, which is analogous to DRP's individual demand contributed at LL as defined in constraint (14).

The price-demand competition is conceived through an interaction mechanism between price and demand using SG. As, in the proposed formulation, each strategic decision variable is affected by the decisions making of other strategic variables due to the common-coupling or shared constraint as defined in (15). This establishes a coupling effect amongst the decision-makers, which deduces a trade-off among DRPs. Hence, it is illustrated through a GSG~\cite{su2004sequential,leyffer2010solving,pang2005quasi}. The equilibrium conditions of the game is termed as GSE~\cite{pang2005quasi}.

In this part, GSG problem is formally presented in the abstract form using the standard game notation and is expressed as follows: $\Upsilon  = \{ {\Omega _\textbf{D}} \cup G,{\{ {S_d}\} _{d \in {\Omega _\textbf{D}}}},{\{ {U_d}\} _{d \in {\Omega _\textbf{D}}}},u_{c}(x)\}$, where ${\Omega _\textbf{D}}$ is set of players, $S$ is set of strategy space of all the players. It consists set of incentive rate $\boldsymbol{\rho}_t^{INC}$ at UL and flexible demand $\boldsymbol{P}_t^F$  at LL. $U$  represents the utility sets, which describes objectives $Z_{LSE}$ and $Z_{DRP,d}$, respectively. $u_{c}(x)$  defines a shared constraint among DRPs. The aspects of GSG game are described in the following section.
\begin{enumerate}[label=(\roman*)]
	\item 	In UL, DRPs' incentive rates are defined as the decision variables. These act as leaders and optimize the LSE cost in conjunction with DRP costs through a set of incentive price $\rho _{d,t}^{INC}$. 
	\item In LL, individual DRP as the aggregated customers acts as follower. It optimizes the customer bill and dis-utility subject to constraints (14)-(15). The shared-constraint $u_{c}(x)$ is defined in (15). It imposes the limits for the possible DRP participation in DS. 
	\item The objectives $Z_{LSE}$ and $Z_{DRP,d}$ are continuously differentiable over $\boldsymbol{\rho} _t^{INC}$   and $\boldsymbol{P} _t^{F}$ , respectively.  In addition, each objective is convex for every   $\rho _{d,t}^{INC}$ and $P_{d,t}^{F}$, and their strategy spaces are bounded. 
\end{enumerate}

Definition 1: Consider a GSG $\Upsilon  = \{ {\Omega _\textbf{D}} \cup G,{\{ {S_d}\} _{d \in {\Omega _\textbf{D}}}},{\{ {U_d}\} _{d \in {\Omega _\textbf{D}}}},u_{c}(x)\}$ with its set of strategies $(\boldsymbol{\rho} _t^{INC },\boldsymbol{P}_t^{F})$. Here, $\boldsymbol{\rho} _t^{INC} = \{ \rho _{1,t}^{INC},\rho _{2,t}^{INC},....,\rho _{{\Omega _D},t}^{INC}\} $ is a set of incentive rate offers to the various DRPs at the UL and  $\boldsymbol{P}_t^{F} = \{ P_{1,t}^{F},..,P_{d,t}^{F},..,P_{{\Omega _D},t}^{F}\}$ is a set of strategies of induced demand by DRPs at the LL. Then, an optimal strategy set $(\boldsymbol{\rho} _t^{INC ^{*} },\boldsymbol{P}_t^{F^{ *} })$  constitutes GSE~\mbox{\cite{tushar2012economics,pang2005quasi,facchinei2007generalized,su2004sequential}}, if the following set of inequalities hold true. 
\begin{gather}
{Z_{DRP,d}}(\rho _{d,t}^{INC ^{*} },P_{d,t}^{F ^{*} },\boldsymbol{P}_{ - d,t}^{F^{*}}) \le {Z_{DRP,d}}(\rho _{d,t}^{INC^{*}},P_{d,t}^F,\boldsymbol{P}_{ - d,t}^{F^{*}}){\rm{     }}\\
{Z_{LSE}}(\rho _{d,t}^{INC^ {*} },P_{d,t}^{F ^{*} },\boldsymbol{\rho} _{ - d,t}^{INC^{*}}) \le {Z_{LSE}}(\rho _{d,t}^{INC},\boldsymbol{P}_{d,t}^{F ^{*} },\boldsymbol{\rho} _{ - d,t}^{INC^{*}}){\rm{    }}\\
\sum\limits_{d \in {\Omega _D}} {P_{d,t}^{F^{*}}}  \le P_{t}^{\max }~~\forall t
\end{gather}
where, $\boldsymbol{\rho} _{ - d,t}^{INC ^{*} } = {(\boldsymbol{\rho} _{d,t}^{INC^{ *} })_{d \in {\Omega _\textbf{D}}\backslash \{ d\} }}$ and $\boldsymbol{P}_{ - d,t}^{F * } = {(\boldsymbol{P}_{d,t}^{F ^{*} })_{d \in {\Omega _\textbf{D}}\backslash \{ d\} }}$   denote the set of strategies i.e. offered incentive prices and induced demands of all DRPs except $d$th DRP. The aforementioned inequalities (19)-(20) state that the players will not have incentive to deviate unilaterally from their equilibrium points. If the players do change their optimality points, then GSE will not constitute.    
\subsubsection{Existence of GSE}
As in the aforesaid non-cooperative game, mainly two types of constraints are involved. First, individual constraint, which depends upon its own variable (i.e. $\rho _{d,t}^{INC}$ and $P_{d,t}^F$) and second is a shared or joint constraint, those which depend upon on all variables i.e., $\boldsymbol{P}_t^F$. These individual constraints are usually satisfied through a pre-defined set of feasible space. Hence, a shared or coupled constraint is illustrated for the sake of simplicity~\cite{facchinei2007generalized}. The GSE is demonstrated with variational inequalities under a shared constraint. It is a refinement over GSE, which ensures that the existence of GSE will lead to the existence of a variational equilibrium (VE)~\cite{facchinei2007generalized,kulkarni2014shared}. Though, it is not always held in general. Hence, GSE is obtained by solving variational inequalities to ensure the solution of GSE~\cite{facchinei2007generalized}. It is denoted by $VI(X,F)$, where $X \equiv P_d^F \subseteq {\mathbb{R}}^{d}$  defines a set and  $F:\mathbb{R}{^d} \to \mathbb{R}{^d}$ is a point-to-point mapping into itself. It states that if $x$ is the solution of variational inequalities, then it will also be the solution of GSE, and GSE should satisfy the following condition: 
\begin{equation}
F{(x)^T}(z - x) \ge 0,{\rm{   }}\forall {\rm{ }}z \in X(x)
\end{equation}where, $x \equiv \{P_{1,t}^{F^{*}},..,P_{d,t}^{F^{*}}\}$ is a solution of variational inequalities and $F(x) \equiv {({\nabla _{{x_1}}}\varphi {(x)^T},....,{\nabla _{{x_2}}}\varphi {(x)^T},...,{\nabla _{{x_{{\Omega _D}}}}}\varphi {(x)^T})^T},{\rm{   }}x \in X$ denotes the partial derivative of each DRP function $\varphi (x) = {Z_{DRP,d}}(P_{d,t}^F,P_{ - d,t}^F)$.

\textbf{Theorem 1:} For an offered incentive rate $\boldsymbol{\rho} _t^{INC}$, a GSE will exist for a game $\Upsilon $ between LSE and DRPs.\\
\textbf{Proof:} In order to find GSE, first DRP problem is reformulated as unconstrained optimization using Lagrange multipliers~\cite{bertsekas2014constrained}. The overall DRPs function with the shared-constraint is formulated as follows:
\begin{equation}
\begin{gathered}
	{Z_{DRP}} = \sum\limits_{d \in {\Omega _D}} {\sum\limits_{t \in {\Omega _T}} {\left\{ {[w_{d,1}(P_{d,t,o}^F - P_{d,t}^F)\rho _{d,t}^{FR} - P_{d,t}^F\rho _{d,t}^{INC}] + w_{d,2}{U_D}(P_{d,t}^F)} \right\}} } \\
	+ {\lambda _d}\left( {\sum\limits_{d \in {\Omega _D}} {P_{d,t}^F}  - P_{t}^{\max }} \right)
\end{gathered}
\end{equation}
Now, the optimality condition for DRP  is obtained through KKT.
\begin{gather}
 - {\nabla _{P_d^F}}{Z_{DRP,d}}(P_{d,t}^F,P_{ - d,t}^F,\rho _{d,t}^{INC}) + {\nabla _{P_d^F}}\left( {\sum\limits_{d \in {\Omega _D}} {P_{d,t}^F}  - P_{t}^{\max }} \right){\lambda _d} = 0\\
0 \le {\lambda _d} \bot \left( {\sum\limits_{d \in {\Omega _D}} {P_{d,t}^F}  - P_{t}^{\max }} \right) \le 0
\end{gather}
where, ${\lambda _d}$ is Lagrange multiplier associated with DRP $d$. The KKT conditions state that the GSE exists for the follower game at a given fixed incentive rate, only if $(P_{d,t}^{F^{*}},{\lambda _d})$ satisfies KKT conditions (24)-(25). Further, the GSE solution is obtained through variational inequalities. It states that if constraint qualification holds at the solution $\boldsymbol{P}_t^F$, then the KKT conditions for the variational inequalities are defined as follows\mbox{\cite{facchinei2007generalized}}:
\begin{gather}
		- {\nabla _{P_d^F}}{Z_{DRP,d}}(P_{d,t}^F,P_{ - d,t}^F,\rho _{d,t}^{INC}) + {\nabla _{P_d^F}}\left( {\sum\limits_{d \in {\Omega _D}} {P_{d,t}^F}  - P_t^{\max }} \right)\lambda  = 0\\
		0 \le \lambda  \bot \left( {\sum\limits_{d \in {\Omega _D}} {P_{d,t}^F}  - P_t^{\max }} \right) \le 0
\end{gather} 

On comparing the KKT conditions (24)-(25) of GSE to the KKT conditions (26)-(27) of variational inequalities $VI(X,F)$, demonstrates that Lagrange multipliers will be common for all DRPs i.e., \mbox{${\lambda _1} = {\lambda _1} = .. = {\lambda _{{\Omega _D}}} = \lambda $}. It indicates that the solution of GSE is variational equilibrium when multipliers values are same for all DRPs\mbox{\cite{facchinei2007generalized,tushar2012economics,li2018interactive}}. This defines the existence of GSE condition. If this GSE condition does not hold, it means that the function is not a convex and not a continuously differentiable\mbox{\cite{facchinei2007generalized}}.
\subsubsection{Uniqueness of GSE}
The uniqueness of GSE determines the social stable outcome of game. It is obtained through the existence of one GSE. It is derived through variational inequalities and a coupling constraints. 

\textbf{Theorem 2:} For the defined game $\Upsilon $, only one GSE exists. 
\textbf{Proof:} In order to verify the uniqueness of GSE, first function $F$ is obtained as follows:  
\begin{equation}
F = \left[ {\begin{array}{*{20}{c}}
		{ - w_{1,1}(\rho _{1,t}^{FR} + \rho _{1,t}^{INC}) +  w_{1,2}({\theta _1}P_{1,t}^F + {\gamma _1})}\\
		{ - w_{2,1}(\rho _{2,t}^{FR} + \rho _{2,t}^{INC}) + w_{2,2}({\theta _2}P_{2,t}^F + {\gamma _2})}\\
		\vdots \\
		{ - w_{d,1}(\rho _{d,t}^{FR} + \rho _{d,t}^{INC}) + w_{d,2}({\theta _d}P_{d,t}^F + {\gamma _d})}
\end{array}} \right]
\end{equation}
Eq. (28) is a partial derivative of DRP function to $P_{d,t}^F$. Now, Jacobian of $F$ gives: 
\begin{equation}
\textbf{J}F = \left[ {\begin{array}{*{20}{c}}
		{{w_{1,2}\theta _1}}&0& \cdots &0\\
		0&{{w_{2,2}\theta _2}}& \cdots &0\\
		0&0& \ddots &0\\
		0&0& \cdots &{{w_{d,2}\theta _d}}
\end{array}} \right]
\end{equation}Eq. (29) is the Jacobian matrix of $F$. It is a positive definite matrix with all positive diagonal elements as ${\theta _d} > 0$. This indicates that $F$ is a strongly monotonic function, which confirms the uniqueness of solution $(\boldsymbol{P}_t^{F^{*}},F(\boldsymbol{P}_t^{F^{*}}))$  of variational inequalities~\mbox{\cite{tushar2012economics,facchinei2007generalized,pang2005quasi}}. It is also equivalent to a unique solution to GSG. This concludes the proof of Theorem 2~\cite{facchinei2007generalized}. 

\vspace{-1mm}
\section{Solution Approach}
The MLMF problem involves multiple Stackelberg problems, which compete non-cooperatively to reach GSE. Mathematically, it is formulated as EPEC to find the equilibrium points, which solves the various MPECs problems simultaneously\mbox{\cite{su2004sequential,hobbs2000strategic,cardell1997market}}. 

\vspace{-1mm}
\subsection{Equivalent NLP Formulation}
In order to illustrate the presented bi-level problem in the equivalent formulation, LL problem is incorporated in UL using MPEC, which is obtained through KKT conditions. The formulation is written compactly, and its subs and superscripts are kept at minimum for the sake of clarity. The UL and LL problem are denoted using the subscripts $u$ and $l$, respectively. The complete formulation of the bi-level problem is as follows: 
\begin{gather}
\mathop {\min }\limits_{{\boldsymbol{x}},{x_d}} {\rm{   }}{Z_{LSE}}({x_d},\boldsymbol{x},x_{ - d}^*,y)\\
{\rm{subject ~~to:~~}}{g_{u,PE}}(\boldsymbol{x},y) = 0{\rm{    }}:{\lambda _P}\\
{g_{u,QE}}(\boldsymbol{x},y) = 0{\rm{    }}:{\lambda _Q}\\
{g_{u,{\rho ^{INC}}}}({x_d}) \ge 0{\rm{    }}:{\boldsymbol{\mu} ^{INC}}\\
{h_{l,E}}({x_d},y,{\boldsymbol{\upsilon} _d}) = {{\nabla} _y}{Z_{DRP}}(y) - {\boldsymbol{\upsilon} _d}{{\nabla} _y}{h_{l,I}}(y) = 0\\
{g_{u,I}}(\boldsymbol{x}) \ge 0{\rm{    }}:\boldsymbol{\pi} \\
0 \le {\boldsymbol{\upsilon} _d} \bot {h_{l,I}}(y) \le 0
\end{gather}
where (34)-(36) are the stationarity conditions of MPECs of LL problem. It is in fact a mathematical program with complementarity constraints (MPCC) due to (36)~\cite{fletcher2004solving}. The UL problem is represented through (30)-(33). LSE' control and state variables such as ${P_{G,n,t}}$, ${Q_{G,n,t}}$, ${e_{n,t}}$ and ${f_{n,t}}$ are confined in $\boldsymbol{x}$ in UL and DRPs variable in UL is denoted by ${x_d} = \{ \rho _{d,t}^{INC}\} $. The individual DRP's decision variable at LL is denoted by $y = \{ P_{d,t}^F\} $. The LSE's active and reactive equality and inequality constraints are comprised into ${g_{u,PE}}(\boldsymbol{x},y)$, ${g_{u,QE}}(\boldsymbol{x},y)$, and ${g_{u,I}}$ respectively in UL. Further, their dual variables associated with active and reactive equality constraints are enclosed in ${\lambda _P} = \{ {\lambda _{p,n,t}}\}$ and ${\lambda _Q} = \{ {\lambda _{q,n,t}}\}$, respectively. LSE’s inequality constraints (8)-(11) are abbreviated in $\boldsymbol{\pi}$. The inequality constraints (14)-(15) of LL are denoted by ${h_{l,I}}(y)$ and their dual variables by $\nu  = \{ {\boldsymbol{\nu} _{{P^F},d,t}},{\underline{\boldsymbol{\nu}}  _{{P^F},d,t}},{\overline{\boldsymbol{\nu}}  _{{P^F},d,t}}\} $. The nonlinear complementarity constraints (36) of LL can be further converted into equality constraints by introducing a slack variable $s$. This formulates the DRP problem in the LL as follows:
\begin{gather}
{h_{l,E}}({x_d},y,{\boldsymbol{\upsilon }_d}) = 0{\rm{    }}:{\phi _{E,d}}\\
{h_{l,I}}(y) - s = 0{\rm{    :}}~{\phi _{I,d}}\\
s \ge 0~~{\rm{    :}}{\phi _{1,d}};~{\boldsymbol{\upsilon} _d} \ge 0~~{\rm{     :}}{\phi _{2,d}}\\
s \circ {\boldsymbol{\upsilon} _d} \le 0{\rm{    :}}~{\phi _{3,d}};~~~~ {\phi _{E,d}},{\phi _{I,d}}~~~~~{\rm{    free}}
\end{gather}
where, constraints (37)-(40) are obtained by transforming the complementarity constraints $0 \le c \bot d \ge 0$ into nonlinear constraints in the form as $c \ge 0,d \ge 0,c \circ d \le 0$ , where $ \circ $  is a notation of Hadamard product.

\vspace{-2mm}
\subsection{Nonlinear Complementarity Formulation } 
As, the formulated problem is non-convex owing to nonlinear constraints and complementarity constraints. Hence, KKT conditions of the problem (30)-(35) and (37)-(40) are strong stationarity conditions rather than the equilibrium conditions~\cite{bautista2007formulation}. This concatenation of strong stationarity conditions results in a complete nonlinear complementarity formulation (NCP) as follows:
\begin{gather}
{\nabla_{\boldsymbol{x}}}{Z_{LSE}} - {\lambda _P}{\nabla _{\boldsymbol{x}}}{g_{u,PE}} - {\lambda _Q}{\nabla _{\boldsymbol{x}}}{g_{u,QE}} - \boldsymbol{\pi} {\nabla _x}{g_{u,I}} = 0\\
{\nabla _{{x_d}}}{Z_{LSE}} - {\boldsymbol{\mu} ^{INC}}{\nabla _{{x_d}}}{g_{u,{\rho ^{INC}}}} - {\phi _{E,d}}{\nabla _{{x_d}}}{h_{l,E}} = 0\\
{\nabla _y}{Z_{LSE}} - {\lambda _P}{\nabla _y}{g_{u,PE}} - {\lambda _Q}{\nabla _y}{g_{u,QE}} - {\phi _{E,d}}{\nabla _y}{h_{l,E}} - {\phi _{I,d}}{\nabla _y}{h_{l,I}} = 0\\
- {\phi _{E,d}}{\nabla _{{\boldsymbol{\upsilon} _d}}}{h_{l,E}} - {\phi _{2,d}} + s \circ {\phi _{3,d}} = 0\\
 - {\phi _{I,d}} - {\phi _{1,d}} + {\boldsymbol{\upsilon} _d} \circ {\phi _{3,d}} = 0\\
{g_{u,PE}}(\boldsymbol{x},y) = 0\\
{g_{u,QE}}(\boldsymbol{x},y) = 0\\
{h_{l,I}}(y) - s = 0{\rm{ }}\\
{h_{l,E}}({x_d},y,{\boldsymbol{\upsilon} _d}) = 0\\
0 \le {g_{u,{\rho ^{INC}}}}({x_d}) \bot {\boldsymbol{\mu} ^{INC}} \ge 0\\
0 \le {g_{u,I}}({x_d}) \bot\boldsymbol{ \pi}  \ge 0\\
0 \le s \bot {\phi _{I,d}} \ge 0;~~~~0 \le {\boldsymbol{\upsilon} _d} \bot {\phi _{2,d}} \ge 0\\
0 \le s \circ {\boldsymbol{\upsilon} _d} \bot {\phi _{3,d}} \ge 0
\end{gather}

The above formulation is a non square-NCP problem as (44), (45) and (48) are not uniquely matched to free variables $y$, $s$, ${\boldsymbol{\upsilon} _d}$ and ${\phi _{l,I}}$. In addition, variables $s$ and $\boldsymbol{\upsilon}_d$ are associated with the multiple inequalities (52) and (53). Therefore, the complementarity constraints (52)-(53) are represented with the different sets of inequalities in (54)-(55)~\cite{leyffer2010solving}.
\begin{gather}
0 \le s \bot {\phi _{I,d}} \ge 0;~~~~{\rm{   }}
0 \le {\boldsymbol{\upsilon} _d} \bot {\phi _{2,d}} \ge 0~~~~{\rm{   }}\forall d\\
0 \le s \bot {\boldsymbol{\upsilon} _d} \ge 0;~~~~ {\phi _{3,d}} \ge 0~~~~{\rm{   }}\forall d
\end{gather}

Further, the complementarity constraints (50)-(51) are re-written in the form of equality constraints by introducing the slack variables ${\omega _d}$ and  ${\varpi _d}$. 
\begin{gather}
{g_{u,{\rho ^{INC}}}}({x_d}) - {\omega _d} = 0;~~~~
0 \le {\boldsymbol{\mu} ^{INC}} \bot {\omega _d} \ge 0\\
{g_{u,I}}({x_d}) - {\varpi _d} = 0;~~~~~0 \le \boldsymbol{\pi}  \bot {\varpi _d} \ge 0
\end{gather}
\subsection{NLP Formulation}
 As, NCP formulation has inherent redundancy problem due to large numbers of multipliers and complementarity constraints~\cite{su2004sequential}. Hence, it is reformulated as a nonlinear problem by transforming complementarity constraints into nonlinear constraints~\cite{su2004sequential}. Further, the complementarity constraints are minimized as an objective function using a penalty method. This gives the complete NLP formulation as follows: 
\begin{equation}
{\mathop{\rm min~}\nolimits} {\rm{  }}{C_{Pen}} = \sum\limits_{d \in {\Omega _D}} {{\phi _{1,d}}s + } {\phi _{2,d}}{\boldsymbol{\upsilon} _d} + s{\boldsymbol{\upsilon} _d} + {\omega _d}{\boldsymbol{\mu }^{INC}} + {\varpi _d}\boldsymbol{\pi} 
\end{equation}
subject to:
\begin{gather}
{\nabla_{\boldsymbol{x}}}{Z_{LSE}} - {\lambda _P}{\nabla _{\boldsymbol{x}}}{g_{u,PE}} - {\lambda _Q}{\nabla _{\boldsymbol{x}}}{g_{u,QE}} - \boldsymbol{\pi} {\nabla _{\boldsymbol{x}}}{g_{u,I}} = 0{\rm{    }}\\
{\nabla _{{x_d}}}{Z_{LSE}} - {\boldsymbol{\mu} ^{INC}}{\nabla _{{x_d}}}{g_{u,{\rho ^{INC}}}} - {\phi _{E,d}}{\nabla _{{x_d}}}{h_{l,E}} = 0{\rm{    }}\\
{\nabla _y}{Z_{LSE}} - {\lambda _P}{\nabla _y}{g_{u,PE}} - {\lambda _Q}{\nabla _y}{g_{u,QE}} - {\phi _{E,d}}{\nabla _y}{h_{l,E}} - {\phi _{I,d}}{\nabla _y}{h_{l,I}} = 0{\rm{    }}\\
 - {\phi _{E,d}}{\nabla _{{\boldsymbol{\upsilon} _d}}}{h_{l,E}} - {\phi _{2,d}} + s \circ {\phi _{3,d}} = 0{\rm{    }}~~~\forall d \in {\Omega _D}\\
- {\phi _{I,d}} - {\phi _{1,d}} + {\boldsymbol{\upsilon} _d} \circ {\phi _{3,d}} = 0{\rm{    }}~~~\forall d \in {\Omega _D}\\
{g_{u,PE}}(\boldsymbol{x},y) = 0\\
{g_{u,QE}}(\boldsymbol{x},y) = 0\\
{h_{l,I}}(y) - s = 0{\rm{ }}\\
{h_{l,E}}({x_d},y,{\boldsymbol{\upsilon} _d}) = 0\\
{g_{u,{\rho ^{INC}}}}({x_d}) - {\omega _d} = 0{\rm{    }}~~~\forall d \in {\Omega _D}\\
{g_{u,I}}({x_d}) - {\varpi _d} = 0{\rm{    }}~~~\forall d \in {\Omega _D}\\
{h_{l,I}}(y) - s = 0{\rm{ }}\\
- {\phi _{I,d}} - {\phi _{1,d}} + {\phi _{3,d}} \circ {\boldsymbol{\upsilon} _d} = 0{\rm{    }}~~~\forall d \in {\Omega _D}\\
 - {\phi _{2,d}} + {\phi _{3,d}} + s \circ {\boldsymbol{\upsilon} _d} = 0{\rm{    }}~~~\forall d \in {\Omega _D}\\
{\boldsymbol{\mu} ^{INC}} \ge 0,\boldsymbol{\pi}  \ge 0,{\omega _d} \ge 0,{\varpi _d} \ge 0\\
s \ge 0,{\boldsymbol{\upsilon} _d} \ge 0,{\phi _{1,d}} \ge 0,{\phi _{2,d}} \ge 0,{\phi _{3,d}} \ge 0{\rm{    }}~~\forall d \in {\Omega _D}
\end{gather}

The reformulated problem (58)-(74) gives the complete NLP model, excluding the complementarity constraints from the problem's constraints. These constraints are moved into the objective function as a penalty function. The feasibility of the problem is evaluated by optimizing the complementarity conditions to zero. Since, the problem is non-convex, optimality is not confirmed through KKT conditions only. Instead, it is ensured through the strong stationarity conditions of the problem. It states that if solution set is a local optimal with ${C_{pen}} = 0$ of the problem, then, the solution set will be strong stationary points. These strong stationary conditions are equivalent to KKT conditions of an NLP problem as proved in~\cite{fletcher2004solving}. The diagonalization method is adopted to solve the formulated EPEC problem~\cite{hobbs2000strategic,su2004sequential}. It is an iterative approach for repetitively solving a sequence of MPECs to reach equilibrium points. A diagonalization approach based on nonlinear Gauss-Seidel method is considered and its procedure is described in Algorithm~\mbox{\ref{your_label}}\mbox{\cite{hu2003mathematical,su2004sequential,hobbs2000strategic}}.
\algsetup{
	linenodelimiter = {)}
}
\begin{algorithm}[t]
	\caption{Procedure of diagonalization method based on Gauss-Seidel approach }\label{your_label}
	\begin{algorithmic}[1]
		\STATE  \textbf{Step 0.}  Initialization: Assuming a starting point $({x_d^{(0)}},{y^{(0)}}) = (x_1^{(0)},x_d^{(d)},....,x_{{\Omega _D}}^{(0)},{y^{(0)}})$  , iteration count $j$ and convergence tolerance $\varepsilon$. .
		\STATE  \textbf{Step 1.}  Loop over MPEC problems: Solving a non-linear MPEC problem for current iteration $j$ $({x_{d}^{(j)}},{y^{(j)}})$  , while keeping constant recently updated information of decision variables of others DRPs i.e. $(x_{ - d}^{(j)},y_{ - d}^{(j)}) = (x_1^{(j + 1)},...,x_{d - 1}^{(j + 1)},x_{d + 1}^{(j)}....,x_{{\Omega _D}}^{(j)},{y^{(0)}})$  . Denoting the current loop solution as $X_d^{(j)} = (x_d^{(j)},{y^{(j)}})$. 
		\STATE  \textbf{Step 2.} Check convergence: if $\left\| {X_d^{(j)} - X_d^{(j - 1)}} \right\| \le \varepsilon $, return $(x_d^{(j)},{y^{(j)}})$, “equilibrium is reached”, otherwise $j=j+1$  and go back to step 1.
	\end{algorithmic}
\end{algorithm}

The Gauss-Seidel method solves a single MPEC at a time, and keeping the other MPECs' recently updated solutions constant. This process continues cyclically and terminates when the difference of solutions of current iteration $j$ and previous iteration $j-1$ lies below the convergence tolerance. 
\vspace{-2mm}
\section{Case Studies}
 The formulated IBDR problem and solution algorithm are investigated on IEEE 33-bus test system~\cite{baran1989network} and Indian 108-bus practical distribution system\mbox{\cite{meena2018optimal}}. 

 \vspace{-2mm}
\subsection{Simulation Data}
The considered systems are sectionalized into five different customer classes. The load buses are discretely specified to the different customer classes. Further, the demand proportions of the customers' classes are set accordingly to demand distribution in the Indian energy market~\cite{share}. S/CF values are chosen discretely based on the practical pricing rates in DS and is referred from~\cite{PANDEY2022107597}. The customers' particulars such as class-wise bus, demand, and SC/F are outlined in Table~\mbox{\ref{tab:cust detail}}~\cite{PANDEY2022107597}.

\begin{table}[bt!]
\vspace{-2mm}
	\caption{Details of customer classes.}
	\centering
	\label{tab:cust detail}
		\setlength{\tabcolsep}{3pt}
	\renewcommand{\arraystretch}{1.4}
			\begin{tabular}{lccccccc}
		\hline
		System  & \multicolumn{3}{c}{IEEE-33   bus}                                                   & \multicolumn{3}{c}{Indian   108-bus}                                                 & \multirow{2}{*}{\begin{tabular}[c]{@{}c@{}}\\S/CF \\ ($\kappa_d$)\end{tabular}} \\ \cline{1-7}
		Classes & Buses & \begin{tabular}[c]{@{}c@{}}Demand\\      (kW)\end{tabular} & \begin{tabular}[c]{@{}c@{}}No. of\\   cust. \end{tabular}& Buses  & \begin{tabular}[c]{@{}c@{}}Demand\\      (kW)\end{tabular} & \begin{tabular}[c]{@{}c@{}}No. of\\   cust. \end{tabular} &                                                                       \\ \hline
		R       & 2-10  & 950                                                        & 121            & 2-43  & 3067                                                       & 386            & -0.2                                                                  \\
		LI      & 19-25 & 1290                                                       & 17             & 55-71  & 4560                                                       & 60             & 0                                                                     \\
		MI      & 26-28 & 180                                                        & 6              & 72-89  & 1476                                                       & 49             & 0.2                                                                   \\
		C       & 11-18 & 555                                                        & 28             & 44-54  & 984                                                        & 51             & 1                                                                     \\
		A       & 29-33 & 740                                                        & 63             & 90-108 & 2045                                                       & 171            & -0.5      \\  \hline                                                         

	\end{tabular}
\end{table}

\begin{figure}[t]
	\centering 
 \vspace{-2mm}
	\includegraphics[width=\columnwidth]{{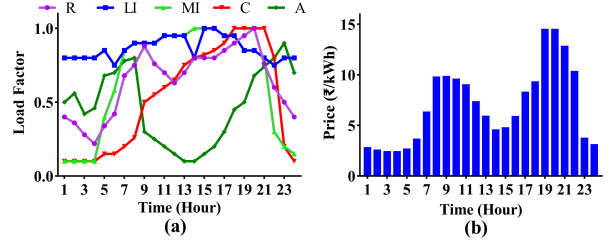}}
        \vspace{-6mm}
	\caption{(a) Class-wise load factor patterns and (b) RTP profile.}\vspace{-3mm}
	\label{load factor}
\end{figure}

The available flexible demand under each DRP is obtained using a truncated normal distribution. Each DRP' flexible demand is the summation of flexible demand under each customer class. The flexible demand is evaluated using the expression ${\varpi _{i,n,d}}\sim \mathcal{N}(\mu ,\sigma );~~{\rm{  }}\varpi _{i,n,d}^{\min } \le {\varpi _{i,n,d}} \le \varpi _{i,n,d}^{\max }$, where  ${\varpi _{i,n,d}}$ is flexible demand factor with their respective bound limits. It is multiplied to the total demand at each node, which gives the flexible demand $P_{n,t}^F$. The DR participation level among the customers is set to be varied within the range of 0-30\%. The class-wise customers' hourly load profiles are obtained using a truncated normal distribution\mbox{\cite{burkardt2014truncated, dadkhah2012cumulant}}. The class-wise representative load profiles are utilized in the normalized form as shown in Fig.~\mbox{\ref{load factor}} (a)\mbox{\cite{kanwar2015optimal}} and represent the Indian Energy market. The complete load profiling modeling may be referred from\mbox{\cite{PANDEY2022107597}}. The RTP prices are referred from the Indian Energy Exchange (INX), as shown in Fig.~\mbox{\ref{load factor}} (b)\mbox{\cite{IndianEnergyExchange}}. The RTP price of December, 2022 are considered for the study and prices are in Indian Indian Rupees (\rupee). The timescale is 24 hours with hourly time resolution in day-ahead operation. Further, the peak periods are considered (8:00-11:00) and (18:00-22:00), respectively to illustrate multi-periods analysis in IBDR. The class-wise offered incentive rates are assumed to be varied within the interval of $(\rho _{d,t}^{\min },\rho _{d,t}^{\max }) \cong (\rho _{d,t}^{FR},\rho _{d,t}^{RTP})$ with the condition of $({\rho _{d,t}} \ge \rho _{d,t}^{FR})$. The dis-utility parameters are set according to~\cite{yu2017incentive}. The contributed flexible demand is taken as a fraction of the total demand. The class-wise retail/FR is set more than the average rate of electricity to hedge against price volatility~\cite{kirschen2000factoring}. It is considered 1.05 times the average of RTP rate in the study. The ratio of \mbox{$\left( {{{{P_{L,n,t}}} \mathord{\left/
			{\vphantom {{{P_{L,n,t}}} {{Q_{L,n,t}}}}} \right.
			\kern-\nulldelimiterspace} {{Q_{L,n,t}}}}} \right)$} is assumed to be the same before DR (BDR) and (after DR) ADR states. Hence, the reactive power will also change accordingly to the active power changes under the IBDR
application.

\begin{figure}[bt!]
	\centering 
 \vspace{-1mm}
	\includegraphics[scale=0.55]{{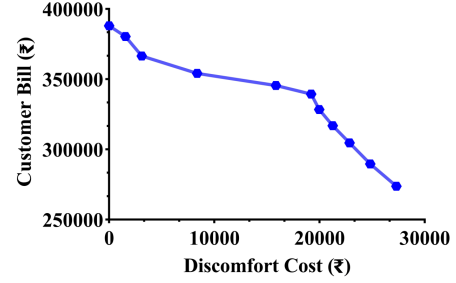}}
        \vspace{-2mm}
	\caption{ Variations in the customer bill and discomfort cost using weighted sum method.}
 \vspace{-5mm}
	\label{weighted-Sum}
\end{figure}

Fig.~\mbox{\ref{weighted-Sum}} illustrates the variations in the customer bill and discomfort cost using the weighted sum method. It is obtained by changing the weights $w_{d,1}$ and $w_{d,2}$ in opposite direction with the step size of 0.1. The figure demonstrates that customer bill is maximum and discomfort cost is minimum when weights are set to at the boundary values (i.e., $w_{d,1}=0$ and $w_{d,2}=1$) and conversely. It indicates zero demand curtailment so zero discomfort cost. However, the customer bill exhibits decrement as $w_{d,1}$ increases and $w_{d,2}$ decreases. 
	Finally, bill reach its minimum value but discomfort cost also become maximum.
The subsequent analyses are performed for $(w_{d,1},w_{d,2})\approx (0.5,0.5)$. 
\vspace{-1mm}
\subsection{IEEE-33 Bus System}
It is rated at 12.66 kV with total active and reactive demands of 3.715 MW and 2.30 MVAR, respectively. The DRPs' locations are pre-established based on the optimal siting of embedded DGs and ESS~\cite{meena2018optimal}. For DR analysis, it is taken as 4, 14, 24, 28, and 30, respectively. Now, the schedules of DRPs and incentive price in the different peak periods obtained from the optimization are summarized in Table~\ref{tab: Optimal incentive price and demand }. The results demonstrate that each customer class’s DRP effectively induces demand curtailment during peak periods based on their incentive rate offers. It is worth to point out that the class-wise incentive rates are obtained after taking into account co-existing demand diversity and cross-subsidy among the different customer classes~\cite{faruqui2012ethics}. This makes the incentive rates to be appeared distinct and different values. If Eq. (12) is not considered, then the incentive rates will be in close proximity for all the customer classes with discrete magnitude in the different time periods. The optimized incentive rate offers for the respective DRPs under the different periods is appeared to be close to retail rate/FR. The variation in incentive rate is highest during peak periods of $\rm{19}^{\rm{th}}$ and $\rm{20}^{\rm{th}}$. Further, demand curtailment is also varying under the different peak periods. It exhibits significant demand curtailment during peak periods of $\rm{19}^{\rm{th}}$ and $\rm{20}^{\rm{th}}$. It is on the account of highest peak RTP price. This causes the utility to offer incentive price, which is less than peak RTP price for the demand reduction. It subsequently lowers the overall cost of LSE.

\begin{table}[bt!]
	\caption{Schedules of class-wise incentive rates and induced flexible demand in the different peak periods.  }
	\label{tab: Optimal incentive price and demand }
	\setlength{\tabcolsep}{1.5pt}
	\renewcommand{\arraystretch}{1.5}
\begin{tabular}{l|ccccccccc}
		\hline
	Index	& $t_8$     & $t_9$     & $t_{10}$    & $t_{11}$    & $t_{18}$    & $t_{19}$    & $t_{20}$    & $t_{21}$    & $t_{22}$    \\ \hline
	$\rho_{1,t}^{INC}$ & 4.51   & 4.51   & 4.51   & 4.51   & 4.51   & 5.35   & 5.38   & 4.69   & 4.51   \\
		$\rho_{2,t}^{INC}$ & 5.64   & 5.64   & 5.64   & 5.64   & 5.64   & 7.24   & 7.32   & 6.44   & 5.64   \\
	$\rho_{3,t}^{INC}$ & 6.76   & 6.76   & 6.76   & 6.76   & 6.76   & 8.16   & 8.18   & 7.13   & 6.76   \\
		$\rho_{4,t}^{INC}$ & 11.27  & 11.27  & 11.27  & 11.27  & 11.27  & 14.05  & 12.49  & 12.53  & 11.27  \\
		$\rho_{5,t}^{INC}$ & 2.82   & 2.82   & 2.82   & 2.82   & 2.82   & 3.55   & 3.63   & 3.20   & 2.82   \\ \hline
	$P_{1,t}^F$	 & 94.40  & 111.82 & 94.95  & 88.07  & 113.39 & 194.38 & 205.06 & 108.45 & 76.52  \\
	$P_{2,t}^F$	 & 134.95 & 135.90 & 131.97 & 140.28 & 139.36 & 248.74 & 251.98 & 173.80 & 109.77 \\
	$P_{3,t}^F$	 & 24.67  & 24.90  & 24.29  & 26.18  & 25.48  & 38.75  & 41.35  & 27.32  & 8.77   \\
	$P_{4,t}^F$ & 18.66  & 36.36  & 39.49  & 43.89  & 68.87  & 124.60 & 92.28  & 92.45  & 57.51  \\
	$P_{5,t}^F$ & 68.95  & 25.59  & 21.91  & 17.56  & 39.12  & 82.29  & 117.78 & 95.09  & 69.61 \\
	 \hline
\end{tabular}
\vspace{-3mm}
\end{table}

The impact of DRPs participation on active power, LSE cost, and DRP cost is illustrated in Table~\ref{tab: Optimal results BDR and ADR  }. The results indicate that DR causes significant changes in the active power drawn by the utility. It is decreased by 13.34\% on average during the peak periods. Similarly, LSE also faces a decrease in the operating cost with the reduced demand except 8, 9, 10, 11 and 18 hours. The increased LSE cost in the exception hours is owing to higher DRP cost. As, DRP’s incentive rate offers should be equal to or greater than retail or flat rate as defined in (3). Further, the highest decrement in the cost is exhibited during peak periods of $\rm{19}^{\rm{th}}$ and $\rm{20}^{\rm{th}}$. In addition, DRP cost indicates granular variation under the different periods. The DRP cost is also highest during the $\rm{19}^{\rm{th}}$ and $\rm{20}^{\rm{th}}$ of peak periods. It demonstrates that LSE will offer higher incentive value to lower its operating cost.    
\begin{table}[bt!]
	\centering
	\caption{Results in the different peak periods BDR and ADR.  }
		\setlength{\tabcolsep}{1.5pt}
	\renewcommand{\arraystretch}{1.5}
	\label{tab: Optimal results BDR and ADR  }	\begin{tabular}{l|cc|ccc}
		\hline
		\multirow{2}{*}{Index} & \multicolumn{2}{c|}{BDR} & \multicolumn{3}{c}{ADR}                 \\ \cline{2-6}
		& $P_G$ (kW) & LSE cost (\rupee) & $P_G$ (kW) & LSE cost (\rupee) & DRP cost (\rupee) \\ \hline
$t_8$ & 3320.27 & 14718.30 & 2949.14 & 14913.18 & 1757.92 \\
$t_9$ & 3175.00 & 14055.89 & 2817.43 & 14286.08 & 1920.82 \\
$t_{10}$& 2970.84 & 12311.20 & 2637.72 & 12626.66 & 1843.34 \\
$t_{11}$& 2995.21 & 10722.16 & 2658.19 & 11228.16 & 1909.32 \\
$t_ {18}$& 3725.84 & 14723.23 & 3300.76 & 15097.63 & 2355.87 \\
$t_{19}$& 3675.14 & 33643.22 & 2919.75 & 31355.41 & 5200.56 \\
$t_{20}$& 3901.03 & 35841.13 & 3115.04 & 33331.71 & 4866.63 \\
$t_{21}$& 3594.67 & 27043.89 & 3040.26 & 25802.86 & 3285.03 \\
$t_{22}$& 3168.32 & 15873.58 & 2811.45 & 15795.36 & 1867.65\\
\hline
	\end{tabular}
 \vspace{-4mm}
\end{table}

\begin{figure}[bt!]
	\centering
	\hspace*{-0.3cm}\includegraphics[scale=0.38]{{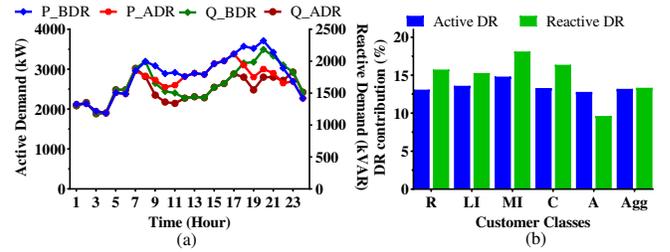}}
 \vspace{-3mm}
	\caption{(a) Load pattern before and after DR and (b) Class-wise demand contribution.}
 \vspace{-4mm}
	\label{agg load patternand DR Con}
\end{figure}

\subsubsection{An aggregated Load Pattern ADR and DR Contribution}
Fig.~\mbox{\ref{agg load patternand DR Con}} (a) illustrates an aggregated active and reactive power load profiles BDR and ADR at the system level. These are obtained by summing up class-wise demand. The figure reveals that the peak active and reactive demand are significantly curtailed during peak periods.
 It is to be noted that the proposed model considers demand curtailment only, as the customers are being compensated through incentives in return. The class-wise DR contributions in terms of active and reactive demands are also depicted in Fig.~\mbox{\ref{agg load patternand DR Con}} (b). It is expressed in terms of the percentage ratios of active and reactive demand variation BDR and ADR. The figure illustrates that each customer class discretely responds in IBDR. The variations in DR contribution across the customers classes are attributed to the different load patterns. This has distinct magnitude of demand at the different hours, which induces distinct demand. The aggregated active DR contribution is obtained as 13.10\%.

\subsubsection{Impact of IBDR on Network Conditions} 
\begin{figure}[bt!]
 \vspace{-3mm}
	\raggedleft
	\hspace*{-0.5cm}\includegraphics[width=0.55\textwidth]{{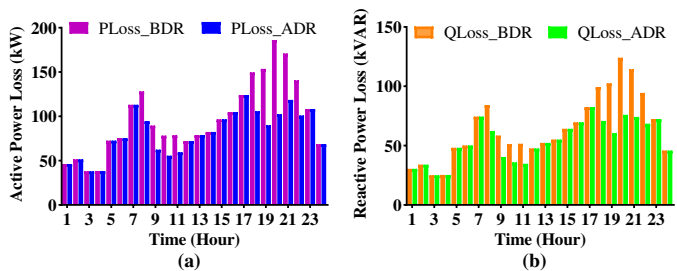}}
	\caption{DR impact on (a) system's active power losses and (b) Reactive power losses. }
 \vspace{-3mm}
	\label{Load factor and power loss V 33}
\end{figure}

The impact of IBDR on the network operating conditions in connection with the active and reactive power losses is shown in Fig.~\mbox{\ref{Load factor and power loss V 33}} (a)-(b). Fig.~\mbox{\ref{Load factor and power loss V 33}} (a) indicates that active power losses are significantly reduced during peak period. Likewise, reactive power losses exhibit decrement during peak periods as observed in Fig.~\mbox{\ref{Load factor and power loss V 33}} (b). The average decrease on active and reactive power losses were found to be around 15.50\% and 15.30\%, respectively.
\subsubsection{Impact of IBDR on Economic Performance}
The impact of the proposed IBDR model on economic performance is described in Table~\ref{tab:Optimization results for a day}. It presents a comparative assessment of the system’s different attributes state under BDR and ADR state over a day. The results indicate that the total energy consumption and LSE cost are lowered by 5.70\% and 2.98\%, respectively. Further, LSE purchasing cost is decreased by 9.45\% ADR. Similarly, the total customer bills, active and reactive power losses are also significantly reduced. These observations illustrate that the incentive offers lower energy requirements, LSE cost, the customer bills, etc. Although, high incentives rates also increase DRP cost. But it would be much better for the LSE to offer the incentives than to pay an exorbitant price when there is no DR during the peak periods. This is confirmed with the reduced LSE cost (2.98\%) and energy consumption (5.70\%) and purchasing cost (9.45\%) after IBDR implementation. Moreover, the class-wise bills BDR and ADR are tabulated in Table~\ref{tab:Class-wise bilss BDR and ADR}. The results exhibit decrement in bills for all the customer classes ADR. It is highest in MI and lowest in LI class customer.

\begin{table}[bt!]
	\centering
	\caption{Application results of the system for a day.}
	\label{tab:Optimization results for a day}
			\setlength{\tabcolsep}{3pt}
	\renewcommand{\arraystretch}{1.3}
		\begin{tabular}{lcccc}
		\hline
Index                          & BDR                   & ADR      & Variation             & \multicolumn{1}{l}{ (\%)} \\ \hline
Total   Energy (kWh)          & 68508.00  & 64600.53  & 3907.48  & 5.70  \\
Avg. Peak   Demand (kW)    & 3261.29   & 2827.12   & 434.16   & 13.31 \\
Flexible   Energy (kWh)       & 7293.40   & 3907.48   & 3385.93  & 46.42 \\
LSE cost   (\rupee/day)           & 150720.39 & 146224.83 & 4495.56  & 2.98  \\
LSE Pur. Cost (\rupee/day) & 536890.24 & 486153.38 & 50736.86 & 9.45  \\
DRP Payment   (\rupee)            & -     & 25007.13  & -        & -     \\
Customer  Bills (\rupee)          & 388071.18 & 340271.00 & 47800.18 & 12.32 \\
Discomfort   Cost (\rupee)            &      -     & 19538.59  &   -       &   -    \\
Active   loss (kWh)      & 2345.72   & 1976.24   & 369.48   & 15.75 \\
 Reactive Loss (KVAR)    & 1556.58   & 1313.44   & 243.14   & 15.62\\  \hline      
\end{tabular}
\end{table}

\begin{table}[bt!]
	\centering
	\caption{Class-wise bills BDR and ADR.}
	\label{tab:Class-wise bilss BDR and ADR}
				\setlength{\tabcolsep}{1.5pt}
	\renewcommand{\arraystretch}{1.3}
		\begin{tabular}{lllllll}
		\hline
Class       & \multicolumn{1}{c}{R  (\rupee)}      & \multicolumn{1}{c}{LI  (\rupee)}     & \multicolumn{1}{c}{MI  (\rupee)}     & \multicolumn{1}{c}{C  (\rupee)}      & \multicolumn{1}{c}{A  (\rupee)}      & \multicolumn{1}{c}{Total Bill  (\rupee)} \\ \hline
BDR         & 76448.05                   & 172590.14                  & 21649.08                   & 89859.37                   & 27524.52                   & 388071.18                      \\
ADR         & 66282.59                   & 155090.55                  & 18256.86                   & 76340.86                   & 24300.14                   & 340271.00                      \\
Change      & -10165.46                  & -17499.60                  & -3392.23                   & -13518.51                  & -3224.38                   & -47800.18                      \\
\multicolumn{1}{c}{(\%) } & \multicolumn{1}{c}{-13.30} & \multicolumn{1}{c}{-10.14} & \multicolumn{1}{c}{-15.67} & \multicolumn{1}{c}{-15.04} & \multicolumn{1}{c}{-11.71} & \multicolumn{1}{c}{-12.32}     \\ \hline
	\end{tabular}
\vspace{-2mm}
\end{table}

\vspace{-2mm}
\subsection{Indian 108-Bus System}
The Indian 108-bus system is rated at 11 kV with an active demand of 12.132 MW and reactive power of 9.099 MVAR, respectively. The system and line data are taken from\mbox{\cite{meena2018optimal}}. The DRPs' locations are predetermined based on the optimal allocation of DGs and ESS~\mbox{\cite{meena2018optimal}}. DRPs locations for the different customer classes are as follows: \mbox{${\rm{R}} = \left\{ {21,24,30} \right\}$}, \mbox{${\rm{LI}} = \left\{ {60,63,67} \right\},{\rm{MI}} = \left\{ {85,88} \right\},{\rm{C}} = \left\{ {45,49} \right\}$} and \mbox{${\rm{A}} = \left\{ {97,102} \right\}$}. The incentive rates and DRPs' induced demands for the different peak periods are summarized in Table~\mbox{\ref{tab: Ch6_optimal price schedule 108}} and Table~\mbox{\ref{tab: Ch6_optimal demand schedule 108}}. The results indicate that the customer classes effectively exhibit the load shaving based on the offered incentive rates. It demonstrates the higher incentive rates and higher demand curtailment as peak periods prices rises. In addition, DRPs of the same customer class exhibit discrete incentive rates and induced demands during the highest peak price periods. 
\begin{table}[!bt]
	\centering
	\renewcommand{\arraystretch}{1.2}
	\setlength\tabcolsep{3pt}
	\caption{Schedules of class-wise incentive rates during the different peak periods}
	\label{tab: Ch6_optimal price schedule 108}
	\begin{tabular}{cccccccccc}
		\hline
		Index              & $t_8$                            & $t_9$                            & $t_{10}$                           & $t_{11}$                           & $t_{18}$                           & $t_{19}$                           & $t_{20}$                           & $t_{21}$                           & $t_{21}$                           \\ \hline
		  & 4.51  & 4.51  & 4.51  & 4.51  & 4.51  & 5.45  & 5.46  & 4.73  & 4.51 \\
		& 4.51  & 4.51  & 4.51  & 4.51  & 4.51  & 5.49  & 5.50  & 4.77  & 4.51  \\ 
		\multirow{-3}{*}{$\rho_{1,t}^{INC}$} & 4.51  & 4.51  & 4.51  & 4.51  & 4.51  & 5.63  & 5.64  & 4.90  & 4.51  \\ \hline
		& 5.64  & 5.64  & 5.64  & 5.64  & 5.64  & 6.86  & 6.87  & 5.97  & 5.64  \\
		& 5.64  & 5.64  & 5.64  & 5.64  & 5.64  & 6.91  & 6.92  & 6.01  & 5.64  \\ 
		\multirow{-3}{*}{$\rho_{2,t}^{INC}$ } & 5.64  & 5.64  & 5.64  & 5.64  & 5.64  & 7.48  & 7.50  & 6.53  & 5.64  \\ \hline
		& 6.76  & 6.76  & 6.76  & 6.76  & 6.76  & 6.76  & 6.76  & 6.76  & 6.76  \\ 
		\multirow{-2}{*}{$\rho_{3,t}^{INC}$} & 6.76  & 6.76  & 6.76  & 6.76  & 6.76  & 6.76  & 6.76  & 6.76  & 6.76  \\ \hline
		& 11.27 & 11.27 & 11.27 & 11.27 & 11.27 & 11.27 & 11.27 & 11.27 & 11.27 \\ 
		\multirow{-2}{*}{$\rho_{4,t}^{INC}$} & 11.27 & 11.27 & 11.27 & 11.27 & 11.27 & 11.27 & 11.27 & 11.27 & 11.27 \\ \hline
		& 2.82  & 2.82  & 2.82  & 2.82  & 2.82  & 3.46  & 3.23  & 3.16  & 2.82  \\
		\multirow{-2}{*}{$\rho_{5,t}^{INC}$} & 2.82  & 2.82  & 2.82  & 2.82  & 2.82  & 3.60  & 3.76  & 3.37  & 2.82  \\
		\hline
	\end{tabular}
\end{table}
\begin{table}[!bt]
	\centering
	\renewcommand{\arraystretch}{1.3}
	\setlength\tabcolsep{1.4pt}
	\caption{Schedules of induced demand during the different peak periods.}
	\label{tab: Ch6_optimal demand schedule 108}
	\begin{tabular}{cccccccccc}
		\hline
		Index              & $t_8$                            & $t_9$                            & $t_{10}$                           & $t_{11}$                           & $t_{18}$                           & $t_{19}$                           & $t_{20}$                           & $t_{21}$                           & $t_{21}$                           \\ \hline
		    & 97.31  & 112.98 & 99.33  & 90.97  & 116.67 & 209.52 & 218.55 & 115.04 & 77.79  \\
		& 97.31  & 112.98 & 99.33  & 90.97  & 116.67 & 213.32 & 222.55 & 117.97 & 77.79  \\
		\multirow{-3}{*}{$P_{1,t}^{F}$} & 97.31  & 112.98 & 99.33  & 90.97  & 116.67 & 226.29 & 236.38 & 126.74 & 77.79  \\ \hline
		& 161.54 & 162.14 & 159.57 & 168.46 & 168.92 & 259.91 & 264.47 & 170.76 & 131.92 \\
		& 161.54 & 162.14 & 159.57 & 168.46 & 168.92 & 264.38 & 269.18 & 174.70 & 131.92 \\
		\multirow{-3}{*}{$P_{2,t}^{F}$}  & 161.54 & 162.14 & 159.57 & 168.46 & 168.92 & 315.50 & 321.61 & 218.60 & 131.92 \\ \hline
		& 66.40  & 66.50  & 66.71  & 36.44  & 66.72  & 82.84  & 66.95  & 66.81  & 31.19  \\
		\multirow{-2}{*}{$P_{3,t}^{F}$}  & 66.51  & 66.00  & 66.71  & 34.78  & 66.72  & 82.84  & 66.87  & 69.26  & 31.19  \\ \hline
		& 15.51  & 29.81  & 33.46  & 35.81  & 46.79  & 46.79  & 46.79  & 47.49  & 46.79  \\
		\multirow{-2}{*}{$P_{4,t}^{F}$}  & 15.51  & 29.81  & 33.46  & 35.81  & 46.79  & 46.79  & 46.79  & 48.10  & 46.79  \\ \hline
		& 105.69 & 40.65  & 33.06  & 26.95  & 61.05  & 117.91 & 134.70 & 138.36 & 106.60 \\
		\multirow{-2}{*}{$P_{1,t}^{F}$}  & 105.69 & 40.65  & 33.06  & 26.95  & 61.05  & 129.49 & 191.51 & 163.99 & 106.60 \\ \hline
	\end{tabular}
\vspace{-2mm}
\end{table}
\subsubsection{An aggregated Load Pattern ADR and DR Contribution}
\begin{figure}[bt!]
	\centering
	\hspace*{-0.3cm}\includegraphics[scale=0.38]{{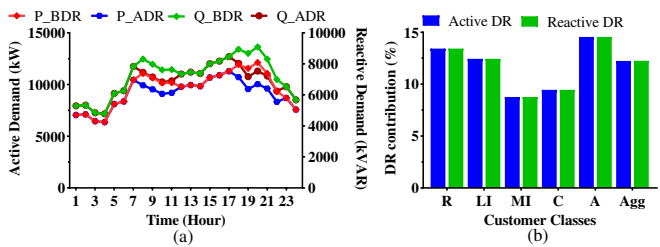}}
 \vspace{-3mm}
	\caption{(a) Load pattern before and after DR and (b) Class-wise demand contribution.}
 \vspace{-3mm}
	\label{agg load patternand DR Con_108}
\end{figure}

The impact of IBDR on the system's active and reactive load profiles are illustrated in Fig.~\mbox{\ref{agg load patternand DR Con_108}} (a). The figure reveals a noteworthy reduction in the demand during peak periods. The class-wise DRPs participation is also illustrated in Fig.~\mbox{\ref{agg load patternand DR Con_108}} (b). The figure demonstrates that all customer classes discretely respond in IBDR. In addition, the active and reactive DR contributions are obtained to be 12.25\% and 12.10\%, respectively.
\subsubsection{Effect of IBDR on Network Operation Conditions}
\begin{figure}[bt!]
	\centering
	\includegraphics[width=0.5\textwidth]{{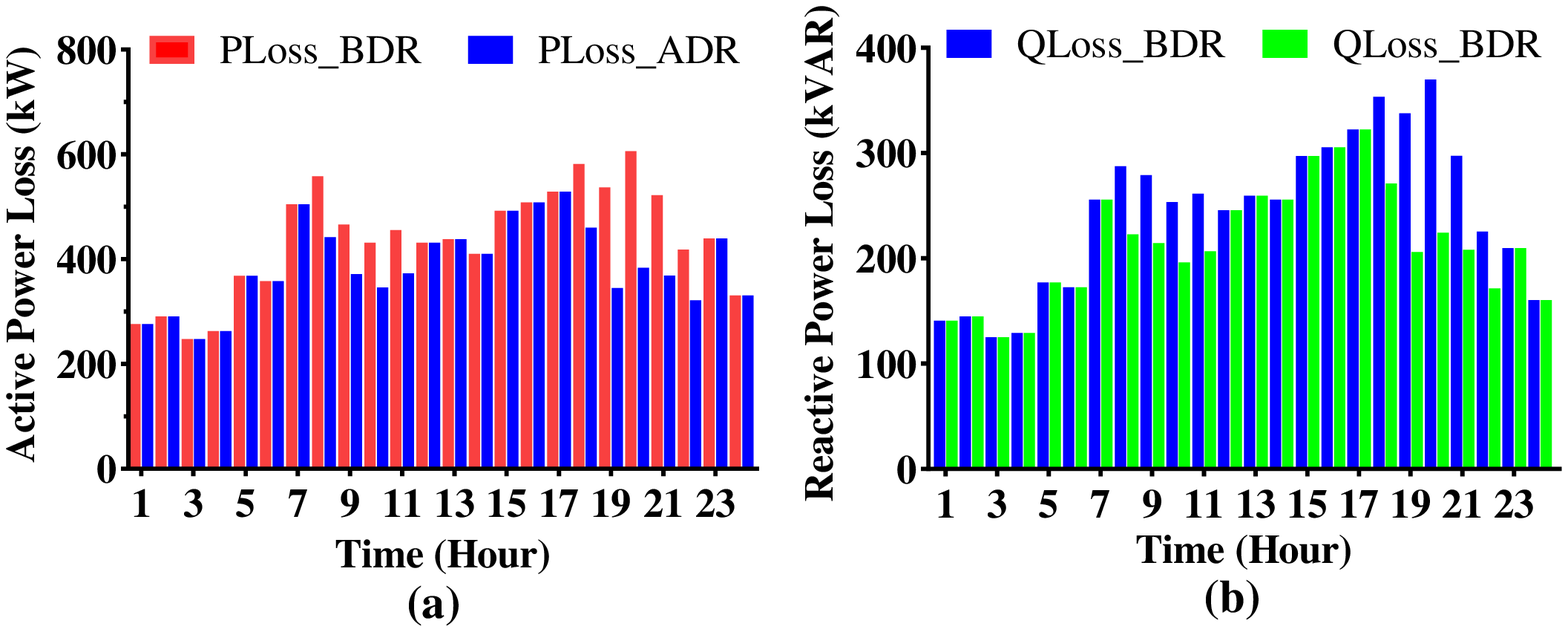}}
 \vspace{-4mm}
	\caption{ DR impact on (a) system's active power losses and (b) Reactive power losses. }
 \vspace{-3mm}
	\label{active and reactive power loss 108}
\end{figure}

The impact of DRPs on active and reactive power losses is exemplified in Fig.~\mbox{\ref{active and reactive power loss 108}} (a)-(b). Both figures demonstrate the reduction of 10.87\% and 12.14\% in the active and reactive losses, respectively.
\subsubsection{Effect of IBDR on Economic Performance}
The effect of the proposed IBDR decision model on the techno-economic performances is described in Table~\mbox{\ref{tab:Ch6_Optimization results 108}}. The results present the variations in the system's different parameters ADR relative to BDR status over a day. The table indicates that the total energy consumption and LSE operating cost are reduced by 5.20\% and 2.64\%, respectively. Hence, this lowers the LSE transaction cost from the grid by 8.52\%, active power losses by 11.13\% and reactive losses 12.67\% respectively, ADR.

\begin{table}[!bt]
	\centering
	\renewcommand{\arraystretch}{1.3}
	\setlength\tabcolsep{1pt}
	\caption{Optimization results of the system for a day}
	\label{tab:Ch6_Optimization results 108}
	\begin{tabular}{lcccc}
		\hline
		Index                       & BDR         & ADR        & Variation & Variation (\%) \\ \hline
		Total   Energy (kWh)           & 230804.96             & 218792.16  & 12012.80              & 5.20                  \\
		Avg. Peak   Demand (kW)     & 10899.35              & 9564.59    & 1334.76               & 12.25                 \\
		Flexible   Energy (kWh)        & 24178.70              & 12012.80   & 12165.90              & 50.32                 \\
		LSE cost   (\rupee/day)            & 515525.48             & 501901.78  & 13623.71              & 2.64                  \\
		LSE pur. cost(\rupee/day) & 1816540.11            & 1661732.15 & 1548.08               & 8.52                  \\
		DRP payment   (\rupee)             & \multicolumn{1}{c}{-} & 69244.38   & \multicolumn{1}{c}{-} & \multicolumn{1}{c}{-} \\
		Customer  Bills (\rupee)           & 1273190.57            & 1139539.79 & 133650.78             & 10.50                 \\
		Disutility   cost (\rupee)         & \multicolumn{1}{c}{-} & 48192.94   & \multicolumn{1}{c}{-} & \multicolumn{1}{c}{-} \\
		Active   loss (kWh)       & 10469.39              & 9304.46    & 1164.92               & 11.13                 \\
		Reactive   loss (kVAR)   & 5868.34               & 5124.59    & 743.74                & 12.67       \\    \hline
	\end{tabular}
 \vspace{-1mm}
\end{table}

\vspace{-1mm}
\subsection{Comparison of the Methods}
A comparison of the solution methods is performed in this subsection. A sample result of class-wise incentive rate offers and induced demand curtailment of  \mbox{$19^{\rm{th}}$} hour for 33-bus systems is presented in Table~\mbox{\ref{tab:Comp optimal schedule}}. The results indicate that the proposed method generates the schedule equivalent to the conventional EPEC approach with negligible error.
\begin{table}[bt!]
	\centering
	\caption{Comparison of schedule using methods.}
	\setlength{\tabcolsep}{4pt}
	\renewcommand{\arraystretch}{1.4}
	\label{tab:Comp optimal schedule}
	\begin{tabular}{lcc|clc}
		\hline
		Method & \multicolumn{2}{c|}{NLP} & \multicolumn{2}{c}{Diagonalization} & \multirow{3}{*}{\begin{tabular}[c]{@{}c@{}}Error\\ (\%)\end{tabular}}\\ \cline{1-5}
		\multirow{2}{*}{Class}  &     $\rho _{d,t}^{INC}$        &     $P_{d,t}^{INC}$        &         $\rho _{d,t}^{INC}$        &         $P _{d,t}^{INC}$        \\
		& (\rupee/kWh)&(kW) &(\rupee/kWh)  & (kW) \\ \hline
	R  & 5.36  & 194.59 & 5.35  & 194.38 & 0.062/0.110  \\
	LI & 7.25  & 249.38 & 7.24  & 248.74 & 0.137/0.260  \\
	MI & 8.16  & 38.77  & 8.16  & 38.75  & 0.038/0.050  \\
	C  & 14.05 & 124.59 & 14.05 & 124.60 & 0.007/-0.005 \\
	A  & 3.56  & 82.10  & 3.55  & 82.29  & 0.243/-0.227 \\\hline        
	\end{tabular}
 \vspace{-3mm}
\end{table}

In addition, the economic comparisons of the results using convex power flow (CPF), NLP, and diagonalization (Diag) are summarized in Table~\mbox{\ref{tab: Eco Comp 33 and 108-bus}}. The CPF is formulated using second-order conic formulation\mbox{\cite{kocuk2017new}}. The results indicate that CLP, NLP and Diag yield comparable results on both case studies. In addition, the comparisons demonstrate equivalent solutions with small variations on the small to large-scale systems. Further, it is worth noting that that NLP and CPF are based EPEC approach, which solve the problem in the centralized way knowing that all decision makers' information available to LSE for optimization. On the other hand, Diag solves the problem in the decentralized way, which handles one decision maker problem at a time through MPEC approach. This enables Diag approach to efficiently describe the strategic behavior of decision makers for imitating non-cooperative behavior based on game theory. 
\begin{table*}[bt!]
		\centering
	\renewcommand{\arraystretch}{1.3}
	\setlength\tabcolsep{6.0pt}
	\caption{Economic comparison for a day using the different methods.}
	\label{tab: Eco Comp 33 and 108-bus}
	\begin{tabular}{lllllll}
		\hline
		System                     & \multicolumn{3}{c}{IEEE-33 bus} & \multicolumn{3}{c}{Indian-108 bus} \\ \hline
		Index                      & CPF       & NLP      & Diag     & CPF        & NLP       & Diag      \\
		Total Energy (kWh)         & 64605     & 64605    & 64601    & 218828     & 217969    & 218792    \\
		Flexible Energy (kWh)      & 3875.20   & 3874.95  & 3907.48  & 11977.26   & 12835.70  & 12012.80  \\
		DR Contribution Rate (\%)      & 13.21     & 13.21    & 13.31    & 12.21      & 13.09     & 12.25     \\
		LSE Cost (\rupee k/day)          & 146.19    & 146.19   & 146.22   & 502.36     & 505.79    & 501.90    \\
		DRP Cost (\rupee k)              & 24.41     & 24.41    & 25.01    & 69.51      & 70.14     & 69.24     \\
		Customer    Bills (\rupee k)     & 341.08    & 341.08   & 340.27   & 1139.00    & 1127.44   & 1139.54   \\
		Total Power Loss (kWh)     & 1977.11   & 1977.13  & 1976.24  & 9315.81    & 9256.08   & 9304.46   \\
		Total Reactive Loss (KVAR) & 1313.96   & 1313.97  & 1313.44  & 5127.63    & 5085.58   & 5124.59  \\ \hline
	\end{tabular}
 \vspace{-2mm}
\end{table*}

A comparison of IBDR performance under the different DR incentive rate structures and different forms of pricing is summarized in Table~\mbox{\ref{tab: Comp_diff_Case}}. The different cases are described as follows: Case 0: BDR, Case 1/2/3: Constrained incentive rate+FR/TOU/RTP, Case 4/5/6: Unconstrained incentive rate+FR/TOU/RTP, and Case 7/8/9: Upper incentive rate limit+FR/TOU/RTP. 
The results indicate that DR participation decreases as the pricing rate changes from FR to TOU or RTP for the constrained incentive rate in case 1, 2, 3. For unconstrained incentive rate, DR participation is found to be negligible, whereas LSE and customers bill exhibit variations on account of the different pricing in cases 4, 5 \& 6. For the upper incentive rate limits in case 7, 8 \& 9, DR manifests participation, but it is lesser than the constrained incentive rates. This demonstrates that incentive rates should be constrained and should have lower threshold limit for inducing appropriate DR, otherwise the customer will participate moderately. Most importantly, the comparison among the different cases indicates that IBDR performs better on the static (i.e., FR) and moderate pricing (i.e, TOU) than highly dynamic pricing (i.e., RTP).

\begin{table}[bt!]
	\centering
	\caption{Comparison of IBDR performance under different cases for IEEE-33 bus system.}
	\label{tab: Comp_diff_Case}
		\renewcommand{\arraystretch}{1.3}
	\setlength\tabcolsep{4pt}
	\begin{tabular}{lccccc}
		\hline
		Case & \begin{tabular}[c]{@{}c@{}}Total Energy \\ (kWh)\end{tabular} & \begin{tabular}[c]{@{}c@{}}LSE Cost\\ (\rupee  k/day)\end{tabular} & \begin{tabular}[c]{@{}c@{}}Customer \\Bill (\rupee k/day)\end{tabular} & \begin{tabular}[c]{@{}c@{}}Avg. Peak   \\  Demand (kW)\end{tabular} & \begin{tabular}[c]{@{}c@{}}DR\\ (\%)\end{tabular} \\ \hline
0 & 68480 & 150.61 & 387.92 & 3259.97 & 0     \\
1 & 64605 & 146.19 & 341.08 & 2829.42 & 13.21 \\
2 & 64907 & 93.79  & 568.22 & 2862.95 & 12.18 \\
3 & 64810 & 142.14 & 560.89 & 2852.18 & 11.51 \\
4 & 68444 & 150.61 & 387.48 & 3255.89 & 0.13  \\
5 & 68454 & 143.06 & 485.09 & 3257.04 & 0.09  \\
6 & 68457 & 145.87 & 635.85 & 3257.38 & 0.08  \\
7 & 65504 & 138.87 & 360.33 & 2929.25 & 10.15 \\
8 & 66340 & 84.19  & 599.90 & 3022.10 & 7.3   \\
9 & 65557 & 133.80 & 583.45 & 2980.17 & 7.01  \\ \hline                                      
	\end{tabular}
\end{table}

\vspace{-2mm}
\subsection{Computational Performance }
The proposed EPEC model and methods are implemented using MATLAB\textsuperscript{\textregistered} toolbox YALMIP~\mbox{\cite{lofberg2004yalmip}} and IPOPT solver~\mbox{\cite{wachter2006implementation}} on Window 10 based personal computer Intel(R) Core(TM) i3-8130U CPU @ 2.20GHz, 4GB RAM. The convergence tolerance value for the diagonalization method is set to 0.01. It converges within 2-3 iterations for the considered system.  The computational performances for the considered test systems via both approaches are presented in Table~\mbox{\ref{tab: Ch6_Compu com}}. The result indicates that the computational costs of the algorithms have a notable time difference for the system under both methods. The diagonalization method produces a smaller size per problem, which lowers constraint enumeration per iteration. Further, it is simple to solve the problem. Though the diagonalization method solves the problem faster, its cycling process may take longer to converge under certain circumstances {\cite{haghighat2012bilevel}}.

\begin{table}[!bt]
	\centering
		\renewcommand{\arraystretch}{1.3}
	\setlength\tabcolsep{2pt}
	\caption{Computational comparison of the methods.}
	\label{tab: Ch6_Compu com}
	\begin{tabular}{lcccc}
		\hline
		System                          & \multicolumn{2}{c}{Standard 33-bus}        & \multicolumn{2}{c}{Indian 108-bus}         \\ \hline
		Method                          & \multicolumn{1}{c}{NLP}  & Diag & \multicolumn{1}{c}{NLP}  & Diag \\ \hline
		\# Variables in UL              & \multicolumn{1}{c}{657}  & 621             & \multicolumn{1}{c}{2043} & 1971            \\ 
		\# Variables in LL              & \multicolumn{1}{c}{189}  & 45              & \multicolumn{1}{c}{333}  & 45              \\ 
		\# Equality constraints in UL   & \multicolumn{1}{c}{612}  & 612             & \multicolumn{1}{c}{1962} & 1962            \\ 
		\# Inequality constraints in UL & \multicolumn{1}{c}{1035} & 918             & \multicolumn{1}{c}{3168} & 2952            \\ 
		\# Equality constraints in LL   & \multicolumn{1}{c}{45}   & 9               & \multicolumn{1}{c}{81}   & 9               \\ 
		\# Inequality constraints in LL & \multicolumn{1}{c}{270}  & 54              & \multicolumn{1}{c}{486}  & 54              \\ 
		CPU Time (sec)                  & \multicolumn{1}{c}{78}   & 14              & \multicolumn{1}{c}{988}  & 489             \\ \hline
	\end{tabular}
\end{table}

\vspace{-2mm}
\section{Conclusions}
A bi-level IBDR decision framework considering multiple DRPs is presented in DS. It is formulated in a competitive environment, where multiple DRPs as DR retail stakeholders compete over incentive rates to improve the LSE's economic performance at the UL and is constrained by individual DRP's objectives at the LL. The interaction among the various DRPs is presented in a non-cooperative environment using a generalized Stackelberg game owing to a multi-leader-multi-follower game. The equilibrium conditions are validated using the variational inequalities. The problem is mathematically formulated as an EPEC problem and is modeled using a nonlinear programming considering AC network constraints. Numerical results demonstrate the effectiveness of the proposed competitive framework for enumerating the potential benefits of each participating players. In addition, the solution algorithm runs adequately on both systems with a reasonable computation time. The comprehensive analyses reveal a strategic behavior among multiple DRPs and its impact on the LSE's operating costs, power losses, incentive valuation, induced DR, etc. It lowers the LSE's operating cost and purchasing cost on account of reduced peak demand. Further, the proposed model exclusively evinces a retail competition among the different DRPs within DS. Moreover, this framework can be extended in the evolving retail market considering other distributed energy resources such as energy storage  and electric vehicles in DS.

\ifCLASSOPTIONcaptionsoff
  \newpage
\fi

\vspace{-2mm}
\bibliographystyle{IEEEtran}
\bibliography{IEEEreferences}

\end{document}